\pdfoutput=1 

\documentclass[journal,article,accept,pdftex,moreauthors]{Definitions/mdpi} 
\DeclareFontFamily{OT1}{pzc}{}
\DeclareFontShape{OT1}{pzc}{m}{it}%
{<-> s * [1.15] pzcmi7t}{}
\DeclareMathAlphabet{\mathpzc}{OT1}{pzc}{m}{it}

\firstpage{1} 
\pubvolume{1}
\issuenum{1}
\articlenumber{0}
\pubyear{2026}
\copyrightyear{2026}
\externaleditor{Firstname Lastname} 
\datereceived{21 August 2026} 
\daterevised{17 November 2026} 
\dateaccepted{23 November 2026} 
\datepublished{ } 

\Title{Hadronic Light-by-Light Piece of the Muon $g-2$: 
Selected Contributions from Dyson–Schwinger Equations
}

\Author{Khépani Raya\textsuperscript{1,2}\orcidA{} 
}

\AuthorNames{Khépani Raya}

\address{%
$^{1}$ \quad Departamento de Actuaría, Física y Matemáticas, Universidad de las Américas Puebla, 72820, San Andrés Cholula, México. \\
$^{2}$ \quad Department of Integrated Sciences and Center for Advanced Studies in Physics, Mathematics and Computation, University of Huelva, E-21071, Huelva, Spain.
}

\corres{Correspondence: khepani.raya@udlap.mx, khepani.raya@dci.uhu.es}

\abstract{The Dyson–Schwinger equations (DSEs) formalism provides a non-perturbative continuum framework for taming the complexities of quantum chromodynamics, becoming a powerful tool for elucidating hadron structure and dynamics. The ability to connect the infrared and perturbative regimes of QCD has also made the DSE framework increasingly relevant to precision tests of the Standard Model, notably the anomalous magnetic moment of the muon ($g-2$). Herein, we revisit DSE-based studies of the hadronic light-by-light scattering contribution to the muon $g-2$. In particular, we examine the pseudoscalar-box and pseudoscalar-pole contributions, discuss the axial-vector meson counterpart, and present \mbox{a first DSE} exploration of the proton-box contribution. We provide a concise overview of the underlying theoretical framework and highlight the interplay between emergent hadronic phenomena and precision phenomenology.}


\begin{document}

\section{Introduction}
\begin{figure}[H]%
\begin{tabular}{lr}
\includegraphics[width=0.46\textwidth]{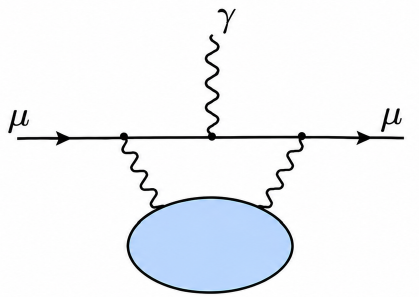} &
\includegraphics[width=0.46\textwidth]{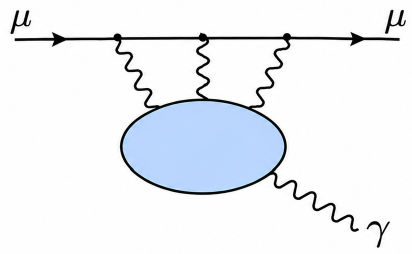}
\end{tabular}
\caption{Hadronic
 contributions to $a_\mu$. (\textbf{Left}) Hadronic vacuum polarization. (\textbf{Right}) Light-by-light scattering. The blue blobs represent the corresponding hadronic interaction subdiagrams.\label{fig:conts}}
\end{figure}

The anomalous magnetic moment of the muon, often referred to as \mbox{muon $g-2$ or $a_\mu$,} continues to stand out as one of the most intriguing observables in modern physics. The excitement surrounding its study has been fueled by the most recent analysis from the Muon $g-2$
 experiment at Fermilab\,\cite{Muong-2:2025xyk}. Based on the latest Fermilab analysis, earlier runs\,\cite{Muong-2:2023cdq,Muong-2:2024hpx,Muong-2:2021ojo,Muong-2:2021vma,Muong-2:2021ovs,Muong-2:2021xzz}, and the historic Brookhaven analysis\,\cite{Muong-2:2006rrc}, the empirical value of $a_\mu$ has been determined as \mbox{$a_\mu^{\text{Exp}}$ = 116,592,071.5(14.5)
 $\times$ $10^{-11}$.} Remarkably, the uncertainty associated with this quantity has been reduced to 124 points per billion. On the theory side, the Muon $g-2$ Theory Initiative has reached a new consensus, establishing $a_\mu^{\text{SM}}$ = 116,592,033(62) $\times$ $10^{-11}$ as the Standard Model (SM) prediction\,\cite{Aliberti:2025beg}. This new assessment emerges from the confluences of numerous analyses (see Refs.\,\cite{Aoyama:2012wk,Volkov:2019phy,Volkov:2024yzc,Aoyama:2024aly,Parker:2018vye,Morel:2020dww,Fan:2022eto,Czarnecki:2002nt,Gnendiger:2013pva,Ludtke:2024ase,Hoferichter:2025yih,RBC:2018dos,Giusti:2019xct,Borsanyi:2020mff,Lehner:2020crt,Wang:2022lkq,Aubin:2022hgm,Ce:2022kxy,ExtendedTwistedMass:2022jpw,RBC:2023pvn,Kuberski:2024bcj,Boccaletti:2024guq,Spiegel:2024dec,RBC:2024fic,Djukanovic:2024cmq,ExtendedTwistedMass:2024nyi,MILC:2024ryz,FermilabLatticeHPQCD:2024ppc,Keshavarzi:2019abf,DiLuzio:2024sps,Kurz:2014wya,Colangelo:2015ama,Masjuan:2017tvw,Colangelo:2017fiz,Hoferichter:2018kwz,Eichmann:2019tjk,Bijnens:2019ghy,Leutgeb:2019gbz,Cappiello:2019hwh,Masjuan:2020jsf,Bijnens:2020xnl,Bijnens:2021jqo,Danilkin:2021icn,Stamen:2022uqh,Leutgeb:2022lqw,Hoferichter:2023tgp,Hoferichter:2024fsj,Estrada:2024cfy,Ludtke:2024ase,Deineka:2024mzt,Eichmann:2024glq,Bijnens:2024jgh,Hoferichter:2024bae,Holz:2024diw,Cappiello:2025fyf,Colangelo:2014qya,Blum:2019ugy,Chao:2021tvp,Chao:2022xzg,Blum:2023vlm,Fodor:2024jyn}) and, in principle, would put an end to the decades-long tension between the SM prediction and the experimental determination of $a_\mu$ (see, e.g., Refs.\,\cite{Aoyama:2020ynm,Colangelo:2022jxc,Jegerlehner:2009ry}). Note, however, that the uncertainty in the SM expectation visibly exceeds its experimental counterpart. While the pure quantum electrodynamics and electroweak contributions are well determined\,\cite{Aoyama:2012wk,Volkov:2019phy,Volkov:2024yzc,Aoyama:2024aly,Parker:2018vye,Morel:2020dww,Fan:2022eto,Czarnecki:2002nt,Gnendiger:2013pva,Ludtke:2024ase,Hoferichter:2025yih}, the hadronic sector remains the main source of uncertainty. This piece is separated into hadronic vacuum polarization (HVP)\,\cite{RBC:2018dos,Giusti:2019xct,Borsanyi:2020mff,Lehner:2020crt,Wang:2022lkq,Aubin:2022hgm,Ce:2022kxy,ExtendedTwistedMass:2022jpw,RBC:2023pvn,Kuberski:2024bcj,Boccaletti:2024guq,Spiegel:2024dec,RBC:2024fic,Djukanovic:2024cmq,ExtendedTwistedMass:2024nyi,MILC:2024ryz,FermilabLatticeHPQCD:2024ppc,Keshavarzi:2019abf,DiLuzio:2024sps,Kurz:2014wya} and hadronic light-by-light (HLbL) contributions\,\,\cite{Colangelo:2015ama,Masjuan:2017tvw,Colangelo:2017fiz,Hoferichter:2018kwz,Eichmann:2019tjk,Bijnens:2019ghy,Leutgeb:2019gbz,Cappiello:2019hwh,Masjuan:2020jsf,Bijnens:2020xnl,Bijnens:2021jqo,Danilkin:2021icn,Stamen:2022uqh,Leutgeb:2022lqw,Hoferichter:2023tgp,Hoferichter:2024fsj,Estrada:2024cfy,Ludtke:2024ase,Deineka:2024mzt,Eichmann:2024glq,Bijnens:2024jgh,Hoferichter:2024bae,Holz:2024diw,Cappiello:2025fyf,Colangelo:2014qya,Blum:2019ugy,Chao:2021tvp,Chao:2022xzg,Blum:2023vlm,Fodor:2024jyn}. The former enters at order $\alpha_{em}^2$, with $\alpha_{em}$ denoting the fine-structure constant in quantum electrodynamics. Its magnitude largely exceeds that of the HLbL piece, which arises at order $\alpha_{em}^3$ but carries the largest relative uncertainty. An illustration of these components is shown in Figure\,\ref{fig:conts}.
 
 \vspace{-9pt}

The intricate nature of the nuclear strong interactions, described within the SM by quantum chromodynamics (QCD), is what prevents the accurate evaluation of the hadronic contributions. QCD is characterized by two key emergent phenomena: the quark–gluon confinement and the dynamical generation of mass in both gauge and matter \mbox{sectors\,\cite{Raya:2024ejx,Roberts:2016vyn,Roberts:2021nhw}.} Confinement prevents the fundamental degrees of freedom, quarks and gluons, from being observed in isolation; instead, experiments can only detect color-singlet composite states known as hadrons. Dynamical mass generation, in turn, is responsible for the emergence of hadron masses (EHM) and thereby accounts for nearly $99\%$ of the mass of visible matter. Both phenomena are intrinsically non-perturbative, therefore inaccessible to perturbative analyses, and are tied to the infrared enhancement of QCD’s process-independent effective charge\,\cite{Brodsky:2010ur,Cui:2019dwv,Deur:2023dzc}. To tame the non-perturbative aspects of QCD, one must thus rely on lattice QCD (lQCD), effective field theories, and other non-perturbative approaches, e.g.,\,\cite{Wilson:1974sk,Weinberg:1978kz,Roberts:1994dr,Brambilla:2014jmp,Estrada:2025zwe}. In the context of the HLbL contribution to $a_\mu$, lQCD\,\cite{Blum:2019ugy,Chao:2021tvp,Chao:2022xzg,Blum:2023vlm,Fodor:2024jyn} and dispersive analyses currently take center stage\,\cite{Colangelo:2015ama,Colangelo:2017fiz,Hoferichter:2018kwz,Bijnens:2019ghy,Bijnens:2020xnl,Bijnens:2021jqo,Danilkin:2021icn,Stamen:2022uqh,Ludtke:2024ase,Deineka:2024mzt,Bijnens:2024jgh,Hoferichter:2024bae,Holz:2024diw}. Even so, complementary data-driven approaches, effective theories and holographic QCD, continue to provide valuable inputs\,\cite{Cappiello:2025fyf,Leutgeb:2022lqw,Cappiello:2019hwh,Leutgeb:2019gbz,Masjuan:2017tvw,Masjuan:2020jsf,Hoferichter:2024fsj,Hoferichter:2023tgp,Estrada:2024cfy, Estrada:2025zwe}. The Dyson–Schwinger equations (DSEs) framework, for its part, is becoming increasingly relevant for addressing the non-perturbative facets of the strong interactions\,\cite{Eichmann:2019tjk,Eichmann:2024glq,Eichmann:2019bqf,Raya:2019dnh,Miramontes:2021exi,Raya:2022ued,Miramontes:2024fgo,Ye:2026zkw,Xing:2025iab,Qin:2019hgk}. Moreover, this framework has proven highly effective in elucidating several properties of many hadrons, covering distribution amplitudes and distribution functions, electromagnetic, electroweak, and gravitational form factors, and complex parton distributions (a rather short representative list includes\,\cite{Raya:2024ejx,Raya:2026gwg,Cui:2020tdf,Roberts:2016vyn,Raya:2021zrz,Roberts:2021nhw,Eichmann:2016yit,Qin:2020rad,Eichmann:2025wgs} and references therein). All of these results arise within a systematically improvable scheme that maintains a traceable connection to QCD.

The present document revisits some of the DSE-based contributions to $a_\mu$ endeavors, with emphasis on the HLbL sector. The manuscript is organized as follows. Section\,\ref{sec:HLbL} offers a concise exposition of the HLbL tensor, capitalizing on the pseudoscalar-pole and box contributions. The presentation of the form factors associated with these components, as well as their expected behavior at low and high energies, is given in Section\,\ref{sec:FFs}. Subsequently, Section~\ref{sec:DSE} provides an overview of the DSE formalism and the evaluation of electromagnetic form factors within this approach. Later on, we present a selection of representative results for {HLbL contributions to $a_\mu$ in Section\,\ref{sec:HLbLDSE} and some perspectives on HVP within DSEs in Section\,\ref{sec:HVP}}.  Finally, a summary and an outlook are given in Section\,\ref{sec:Summary}.

\clearpage 

\section{Hadronic Light-by-Light Tensor}
\label{sec:HLbL}

The HLbL tensor is defined as the hadronic four-point Green’s function of the electromagnetic quark currents
$j_{em}^{\mu}(x)$,
\vspace{-9pt}
\begin{adjustwidth}{-\extralength}{0cm}
\centering 
\begin{equation}
    \Pi^{\mu\nu\lambda\rho}(q_1,q_2,q_3)=-i \int d^4x \,d^4y\,d^4z\, e^{-i(q_1\cdot x+q_2\cdot y+q_3\cdot z)}\langle0|T\{j_{em}^{\mu}(x)j_{em}^{\mu}(y)j_{em}^{\mu}(z)j_{em}^{\mu}(0)\}|0\rangle,
\end{equation}
\end{adjustwidth}
where $q_{1,2,3}$ denote independent timelike photon momenta (the fourth one is fixed by momentum conservation). A consistent set of gauge-invariant Lorentz structures $T_i^{\mu\nu\lambda\sigma}$, accompanied by scalar functions $\Pi_i$ free of kinematic singularities, was derived in Ref.\,\cite{Colangelo:2015ama}. The construction is rooted in the principles established in Refs.\,\cite{Bardeen:1968ebo,Tarrach:1975tu} and yields \mbox{the decomposition:}
\begin{equation}
    \Pi^{\mu\nu\lambda\sigma}=\sum_{i=1}^{54}T_i^{\mu\nu\lambda\sigma}\Pi_i.
\end{equation}

Crossing symmetries, Ward identities, and kinematic relations imply strong linear dependencies among the scalars $\Pi_i$. Therefore, only a much smaller set of independent combinations is required after the appropriate projection\,\cite{Colangelo:2017fiz}. Resulting ambiguities in the definition of the scalar functions, in any case, must cancel out in the eventual calculation of the physical observable $a_\mu$.

The HLbL contribution to $a_\mu$ is then extracted from $\Pi_i^{\mu\nu\lambda\sigma}$ using the standard Dirac-space projection operator techniques of\,\cite{Aldins:1970id}, applied in the limit $q_4\to0$. This implementation yields the well-known master formula:
\begin{equation}
\label{eq:GenHLbL}
   a_{\mu}^{\mathrm{HLbL}}=\frac{\alpha^{3}_{em}}{432\pi^{2}}\int_{0}^{\infty}d\tilde{\Sigma}\,\tilde{\Sigma}^{3}\int_{0}^{1}d r\,r\sqrt{1-r^{2}}\int_{0}^{2r}d\phi\,\sum_{i=1}^{12}T_{i}(Q_{1},Q_{2},Q_3)\tilde{\Pi}_{i}(Q_{1},Q_{2},Q_3),
\end{equation}
where integration variables $(\tilde{\Sigma}, r, \phi)$ parametrize the Euclidean momenta $Q_i^2=-q_i^2$ \mbox{as follows:}
\begin{equation}
   \begin{array}{l}
   Q_{1}^{2}=\frac{{\Sigma}}{3}\left(1-\frac{r}{2}\cos\phi-\frac{r}{2}\sqrt{3}\sin\phi\right),\\ 
   Q_{2}^{2}=\frac{{\Sigma}}{3}\left(1-\frac{r}{2}\cos\phi+\frac{r}{2}\sqrt{3}\sin\phi\right),\\ 
   Q_{3}^{2}=\frac{{\Sigma}}{3}\left(1+r\cos\phi\right).
   \end{array}
\end{equation}

The kernel functions $T_i$ encode the QED and kinematic factors; explicit expressions might be found in \cite{Colangelo:2017fiz} (Appendix B). 
 The functions $\tilde{\Pi}_i$ are properly symmetrized combinations of $\Pi_i$ and therefore contain the dynamical component. In particular, for the meson-box contribution,
\begin{equation}
    \label{eq:PiBox}
    \tilde{\Pi}_i^{\textbf{P}-\text{box}}(Q_1^2,Q_2^2,Q_3^3)=F_\textbf{P}(Q_1^2)F_\textbf{P}(Q_2^2)F_\textbf{P}(Q_3^2)\frac{1}{16\pi^2}\int_0^1 dx\int_0^{1-x}dy,I_i(x,y),
\end{equation}
where the scalar functions $I_i$ are provided in \cite{Colangelo:2017fiz} (Appendix C),
 and $F_\textbf{P}(Q_i^2)$ denotes the $\textbf{P}=\{\pi^+,K^+\}$ elastic electromagnetic form factor (FF). For the pseudoscalar-pole contributions, $\textbf{P}=\{\pi^0,\eta,\eta',\eta_c,\eta_b\}$, it is convenient to express Equation\,\eqref{eq:GenHLbL} in terms of the weight functions $w_{1,2}:=w_{1,2}(Q_1,Q_2,\cos \phi)$, as discussed in, e.g.,\,\cite{Nyffeler:2016gnb,Jegerlehner:2009ry,Roig:2014uja}:
\vspace{-9pt}
\begin{adjustwidth}{-\extralength}{0cm}
\centering 
\begin{equation}
\label{eq:amuPSpole}
    a_\mu^{\textbf{P}-\text{pole}}=\left(\frac{\alpha_{em}}{\pi}\right)^3\int dQ_1 dQ_2 d(\cos\phi)\Big[w_1 G_\textbf{P}(Q_1^2,Q_3^2)G_\textbf{P}(Q_2^2,0)+w_2G_\textbf{P}(Q_1^2,Q_2^2)G_\textbf{P}(Q_3^2,0) \Big];
\end{equation}
\end{adjustwidth}
here, $G_\textbf{P}(Q_i^2,Q_j^2)$ denotes the pseudoscalar meson to two-photon transition FF.

The need for the proper determination of the elastic and transition FFs, $F_\textbf{P}$ and $G_\textbf{P}$, is evident from Equations \eqref{eq:PiBox} and \eqref{eq:amuPSpole}. This necessity is augmented by the dominance of their contributions to the HLbL piece of $a_\mu$. According to the latest consensus\,\cite{Aliberti:2025beg},
\begin{equation}
\label{eq:currAmu}
    a_\mu^{\pi^0,\eta,\eta'-\text{poles}}=91.2_{-2.4}^{+2.9}\times10^
    {-11}\,,\,a_\mu^{\pi,K-\text{box}}=-16.4(2)\times10^{-11}.
\end{equation}

\textls[-5]{Other contributions, such as those from axial-vector and tensor mesons, are numerically subdominant; however, they entail nontrivial theoretical subtleties and sizable uncertainties that require careful assessment (see, e.g., Refs.\,\cite{Roig:2019reh,Mager:2025pvz,Cappiello:2025fyf}). Specific implementations of the master formula, Equation\,\eqref{eq:GenHLbL}, relevant for their evaluation, are presented in
Ref.\,\cite{Hoferichter:2024fsj}.}

\section{Electromagnetic Form Factors}
\label{sec:FFs}

\subsection{Matrix Elements}
Let us consider a pseudoscalar meson composed of a valence quark $q$ and an antiquark $\bar{h}$. When this state is on shell, its interaction with an electromagnetic probe (a photon) is described by a single form factor $F_\textbf{P}(Q^2)$. The corresponding matrix element for this elastic process reads:
\begin{equation}
    \label{eq:MatElemEFF}
    \Lambda_\mu^{\gamma\textbf{P}\textbf{P}}(Q,K)=2K_\mu F_\textbf{P}(Q^2),
\end{equation}
where $Q=p_o-p_i$ is the photon virtuality, $p_i(p_o)$ the incoming (outgoing) meson momentum, and $2K=p_i+p_o$. The two-photon transition to a neutral pseudoscalar meson, $\gamma^*\gamma^*\to \textbf{P}$, is also defined by a single form factor:
\begin{equation}
    \label{eq:MatElemTFF}
    \Lambda_{\mu\nu}^{\gamma\gamma\textbf{P}}(Q_1,Q_2)=e^2\epsilon_{\mu\nu\alpha\beta}Q_{1\alpha}Q_{2\beta}G_\textbf{P}(Q_1^2,Q_2^2),
\end{equation}
with $Q_{1,2}$ denoting the photon virtualities and $e^2=4\pi\alpha_{em}$. Bose symmetry entails $G_\textbf{P}(Q_1^2,Q_2^2)=G_\textbf{P}(Q_2^2,Q_1^2)$.


Notably, experimental data are available for the elastic electromagnetic form factors of both the $\pi^+$\,\cite{NA7:1986vav, Horn:2007ug, JeffersonLab:2008jve} and the $K^+$\,\cite{Dally:1980dj, Amendolia:1986ui, Carmignotto:2018uqj}. Likewise, measurements of the two-photon transition form factors exist for the $\pi^0$\,\cite{CELLO:1990klc,CLEO:1997fho,BaBar:2009rrj,Belle:2012wwz}, $\eta-\eta'$\,\cite{CELLO:1990klc,CLEO:1997fho,L3:1997ocz,BaBar:2011nrp,BaBar:2006ash}, and $\eta_c$\,\cite{BaBar:2010siw}. No experimental data are currently available for the $\eta_b$, but meaningful comparisons can be made with predictions from nonrelativistic QCD\,\cite{Feng:2015uha}.

Electromagnetic FFs not only encode the electromagnetic structure of hadrons, e.g.,\,\cite{Xu:2025sxw,RuizArriola:2025wyq}, but also offer an interpolation between a hadronic description at low momentum transfers and partonic behavior at high energies\,\cite{Farrar:1979aw,Lepage:1979zb,Lepage:1980fj,Efremov:1979qk}. In the following, we discuss the expectations in these energy regimes.

\subsection{Large-$Q^2$ Expectations}
The asymptotic limits of the pion elastic and two-photon transition FFs follow from the factorization properties of QCD expected to hold in the large-momentum \mbox{regime\,\cite{Farrar:1979aw,Lepage:1979zb,Lepage:1980fj,Efremov:1979qk}.} It is within this domain that the expected scaling laws, and their violations, arise. For instance, let us first consider the leading-order expectation for the pseudoscalar meson elastic FF:
\vspace{-3pt}
\begin{align}
Q^2 F_{\textbf{P}}(Q^2) \stackrel{Q^2 \gg m_p^2}{\approx} 16 \pi \alpha_s(Q^2) f_\textbf{P}^2 \mathpzc{w}_\textbf{P}^2(Q^2).
\label{eq:FPUV}
\end{align}

Here, $\alpha_s$ is the one-loop strong running coupling; $f_\textbf{P}$ denotes the meson's leptonic decay constant ($f_\pi \approx 0.092\,$GeV and $f_K\approx0.110\,$GeV); 
and $\mathpzc{w}_\textbf{P}^2 = e_{\bar h} \mathpzc{w}_{\bar h}^2(Q^2) + e_{q}\mathpzc{w}_q^2(Q^2)$, \mbox{such that}
\begin{equation}
\label{weightings}
\mathpzc{w}_{q} = \tfrac{1}{3}\int_0^1 dx\, \mathpzc{g}_q(x) \,\varphi_\textbf{P}^q(x) ,
\end{equation}
$\mathpzc{g}_q(x) = 1/(1-x)$, and $\mathpzc{g}_{\bar h}(x) = 1/x$; $e_{q,\bar{h}}$ denote the valence-quark and antiquark electric charges in units of that of the positron. The quantity $\varphi_{\textbf{P}}^q(x)$ represents the leading-twist valence-quark distribution amplitude (DA), which is expressed in terms of the light-front momentum fraction $x$ and normalized to unity within this convention. At asymptotically large resolving scales, the DA behaves as\,\cite{Farrar:1979aw,Lepage:1979zb,Lepage:1980fj,Efremov:1979qk}
\begin{equation}
    \label{eq:asymDA}
    \varphi^{asy}(x)=6x(1-x).
\end{equation}

In contrast to this profile, the $\pi$ and $K$ DAs at experimentally accessible scales are broad and concave, reflecting the dominant role of EHM in light-quark mass generation. Heavy quarkonia, by comparison, exhibit narrower DAs as Higgs mass generation becomes increasingly important. Heavy-light mesons display a pronounced asymmetry, whereas the much milder asymmetry of the kaon reflects the dominance of EHM. Notably, the $\eta$ and $\eta'$ DAs lie close to the asymptotic form. These systematic trends, and their connection with the emergent facets of QCD, are discussed in detail in Ref.\,\cite{Raya:2024ejx}; in particular, {Figures} 
 \ref{fig:EFFlarge} and \ref{fig:TFFlow} therein provide a clear illustration of all these features.

The leading-order hard-scattering formula for the two-photon TFF reads:
\begin{align}
G_{\textbf{P}}(Q_1^2,Q_2^2) \stackrel{Q_{1,2}^2 \gg m_p^2}{\approx} 2f_\textbf{P}\mathcal{Q}_\textbf{P}^2 \int_0^1dx\,\frac{\varphi_\textbf{P}^q(x)}{x Q_1^2+(1-x)Q_2^2},
\label{eq:GPUV}
\end{align}
with $\mathcal{Q}^2_{\pi,\,\eta_c,\,\eta_b}=\{1/3 ,\,4/9,\,1/9\}$ being a factor arising from the flavor structure of the meson. At asymptotically large energies, where the DA of Equation~\eqref{eq:asymDA} is valid, the following exact asymptotic prediction is obtained\,\cite{Farrar:1979aw,Lepage:1979zb,Lepage:1980fj,Efremov:1979qk}:
\begin{equation}
\label{eq:asymTFF}
    \bar{G}_\textbf{P}^{asy}:=\lim_{Q^2\to\infty}Q^2G_{\textbf{P}}(Q^2,0)\to2f_\textbf{P}\,(3\mathcal{Q}^2).
\end{equation}

A robust DSE treatment naturally incorporates such behavior\,\cite{Raya:2015gva}. Note that for the mixed $\eta-\eta'$ system, Equations\,\eqref{eq:GPUV} and \eqref{eq:asymTFF} must be modified accordingly to account for the flavor structure and the scale evolution of the decay constants\,\cite{Hoferichter:2020lap,Feldmann:1997vc,Agaev:2014wna}. 

The asymptotic expectations from Equations\,\eqref{eq:FPUV} and\,\eqref{eq:GPUV} highlight important features. First, these anticipate a $1/Q^2$ power-law falloff in the asymptotic domain, modified by logarithmic corrections arising from anomalous dimensions. In the elastic case, this is reflected explicitly through the dependence on $\alpha_s$ and, in both FFs, through the scale evolution of the DA\,\cite{Farrar:1979aw,Lepage:1979zb,Lepage:1980fj,Efremov:1979qk}. The decay constant appearing as an overall factor is also noteworthy: Since $f_\textbf{P}$ measures the strength of dynamical chiral symmetry breaking (DCSB), its presence establishes a direct link between the asymptotic domain and the non-perturbative dynamics of QCD.

Over the years, the long-known Equations\,\eqref{eq:FPUV} and \eqref{eq:GPUV} have been refined through the inclusion of higher-order, higher-twist, and explicit mass corrections, e.g.,\,\cite{Hoferichter:2020lap,Chai:2025xuz,Chen:2024oem}. In particular, the treatment of mass effects from\,\cite{Hoferichter:2020lap} aligns with the mechanisms that generate violations of the $x-k_\perp$ factorization of the meson light-front wave function beyond the chiral limit\,\cite{Raya:2021zrz,Raya:2024glv,Albino:2022gzs,Raya:2026gwg,Xu:2026rtu}.

Finally, as a point of reference, the combination of several studies based on a dispersive approach yields\,\cite{Hoferichter:2018kwz,Holz:2024diw,Hoferichter:2018dmo,Holz:2024lom}
\begin{equation}
\label{eq:asydisp}
    \bar{G}_\pi^{asy}=2f_\pi\,,\bar{G}_\eta^{asy}=0.183(13)\,,\bar{G}_{\eta'}^{asy}=0.264(13).
\end{equation}

These quantities will be analyzed from the perspective of the DSE framework in a later section. For the moment, we focus on their low-energy expectations.

\subsection{Low-$Q^2$ Behavior}
In the low-energy limit, gauge invariance enforces the condition $F_\textbf{P}(0)=1$, providing a sensitive check of the consistency of any theoretical framework. For the transition FFs of the Nambu–Goldstone pseudoscalar octet, the corresponding normalization is governed by the so-called chiral anomaly\,\cite{Adler:1969gk,Bell:1969ts}, which yields the chiral-limit constraint:
\begin{equation}
\label{eq:AA}
    G_{\textbf{P}}^0(0,0) = \frac{1}{4\pi^2f_\pi^0}.
\end{equation}

The appearance of $f_\pi^0$, the chiral-limit value of the pion leptonic decay constant, shows that DCSB sets the infrared scale. Deviations from this limit therefore quantify the interplay between EHM and Higgs mass generation. The $\eta-\eta'$ system is special, however, owing to its connection with the non-Abelian axial anomaly. For instance, being excluded from the light-pseudoscalar octet, the $\eta'$ departs substantially from the behavior described by Equation\,\eqref{eq:AA}, even in the chiral limit\,\cite{Raya:2024ejx}.

The $G_\textbf{P}(0,0)$ value can be related to the corresponding decay width $\Gamma_\textbf{P}^{\gamma\gamma}$ via
\begin{equation}
    \bar{G}^0_\textbf{P}:=G_\textbf{P}(0,0)=\sqrt{\frac{4\Gamma_\textbf{P}^{\gamma\gamma}}{\pi \alpha_{em}^2m_\textbf{P}^3}}.
\end{equation}

Employing Particle Data Group (PDG) estimates\,\cite{ParticleDataGroup:2024cfk}, one finds (in units of GeV$^{-1}$)
\begin{eqnarray}
\label{eq:ValuesG0}
    &\bar{G}^0_\pi=0.2739(21)\,,\,\bar{G}^0_\eta=0.2739(48),&\\
    &\bar{G}^0_{\eta'}=0.3437(55)\,,\,\bar{G}^0_{\eta_c}=0.0678(27).&\nonumber
\end{eqnarray}

\textls[-5]{Normalization alone does not exhaust the information contained in the low-momentum region. For instance, in the vicinity of $Q^2\approx0$, the elastic FF admits \mbox{the expansion:}}
\begin{equation}\label{eq:ChargeR}
    F_\textbf{P}(Q^2)=1-\frac{1}{6}r_\textbf{P}^2 Q^2+\cdots\,,\,r_\textbf{P}^2:=-6\frac{\partial F_\textbf{P}(Q^2)}{\partial Q^2}\Bigg|_{Q^2=0},
\end{equation}
where $r_{\textbf{P}}$ is identified as the meson's charge radius. An analogous expression can be derived for the transition form factor,
\begin{equation}
    \label{eq:tranR}
    G_\textbf{P}(Q^2,0)=\bar{G}^0_\textbf{P}[1-b_\textbf{P}Q^2+\cdots]\,,\,b_\textbf{P}:=-\frac{1}{\bar{G}^0_\textbf{P}}\frac{\partial G_\textbf{P}(Q^2,0)}{\partial Q^2}\Bigg|_{Q^2=0}.
\end{equation}

Both $r_\textbf{P}$ and $b_\textbf{P}$ serve as valuable low-energy benchmarks. The charge radius encapsulates the long-distance distribution of the meson electromagnetic current, whereas the slope parameter $b_\textbf{P}$ characterizes the departure from the anomaly-driven normalization of the transition FF. PDG determinations are $r_{\pi^+}=0.659(4)$ fm and $r_{K^+}=0.560(31)$ fm\,\cite{ParticleDataGroup:2024cfk}, while the determination from Ref.\,\cite{Cui:2022fyr} yields $r_{\pi^+}=0.640(7)$ fm. For their part, recent data-driven analyses\,\cite{Hoferichter:2018kwz,Holz:2024diw,Hoferichter:2018dmo,Holz:2024lom} provide the following slope parameters (in GeV$^{-2}$):
\begin{equation}
\label{eq:slopeDD}
    b_{\pi^0}=1.73(5)\,,\,b_\eta=1.83(4)\,,\,b_{\eta'}=1.49(3).
\end{equation}

The fulfillment of the low- and high-energy expectations is crucial to properly capture the transition from the non-perturbative regime (that underlies hadron structure) to the short-distance domain (where partonic behavior emerges). These aspects are essential for a reliable evaluation of the HLbL contributions\,\cite{Melnikov:2003xd,Bijnens:2019ghy,Colangelo:2019uex}. Consequently, \mbox{a framework} capable of reproducing both limits while properly tracing the transition between them is necessitated. These expectations are naturally met within the DSE approach.

\section{The Dyson–Schwinger Equations Framework}
\label{sec:DSE}

\subsection{Dyson–Schwinger and Bethe–Salpeter Equations}
The Dyson–Schwinger equations are the quantum-field-theoretic analogs of the \mbox{Euler–Lagrange} equations. These encode the full dynamics of the underlying theory through an infinite tower of coupled integral equations for its $n$-point Green’s \mbox{functions\,\cite{Roberts:1994dr,Eichmann:2016yit,Qin:2020rad,Eichmann:2025wgs}.} In QCD, for example, the DSE for the quark propagator (\emph{gap equation})
 depends on the gluon two-point function and the quark–gluon three-point vertex. Each of these objects satisfies its own DSE, involving still higher $n$-point functions \emph{ad infinitum}.

To address the bound-state problem within this framework, a natural starting point is therefore the quark gap equation, which reads:
\begin{eqnarray}\label{eq:quarkPropQCD}
    S^{-1}(p)&=&Z_2 [S^{(0)}(p)]^{-1} + \int_{q} \mathcal{K}^{(1)}(q,p) S(q),  \\
    \mathcal{K}^{(1)}(q,p)&=&\frac{4}{3} Z_1 g^2 D_{\mu\nu}(p-q) \gamma_\mu\otimes  \Gamma_\nu(p,q).\nonumber
\end{eqnarray}

Here, $S(p)$ denotes the fully dressed quark propagator and $S^{0}(p)$ its bare counterpart; the 1-body kernel $\mathcal{K}^{(1)}(q,p)$ is defined in terms of the fully dressed gluon propagator, $D_{\mu\nu}(p-q)$, and quark–gluon vertex (QGV), $\Gamma_\nu(p,q)$. The quantities $Z_1$ and $Z_2$ are the renormalization constants associated with the quark–gluon vertex and quark propagator, respectively, and $g$ corresponds to the bare Lagrangian coupling constant. The notation $\int_q:=\int \frac{d^4q}{(2\pi)^4}$ indicates a Poincaré-invariant regularized integral. For simplicity, we have omitted color, flavor and spinor indices.

The dressed-quark propagator may be generically expressed as follows:
\begin{equation}
\label{eq:propGen}
    S(p)=Z(p^2)(i\gamma\cdot p+M(p^2))^{-1}.
\end{equation}

\begin{figure}[H]%

\begin{adjustwidth}{-\extralength}{0cm}
\centering 
\begin{tabular}{lr}
\includegraphics[width=0.6\textwidth]{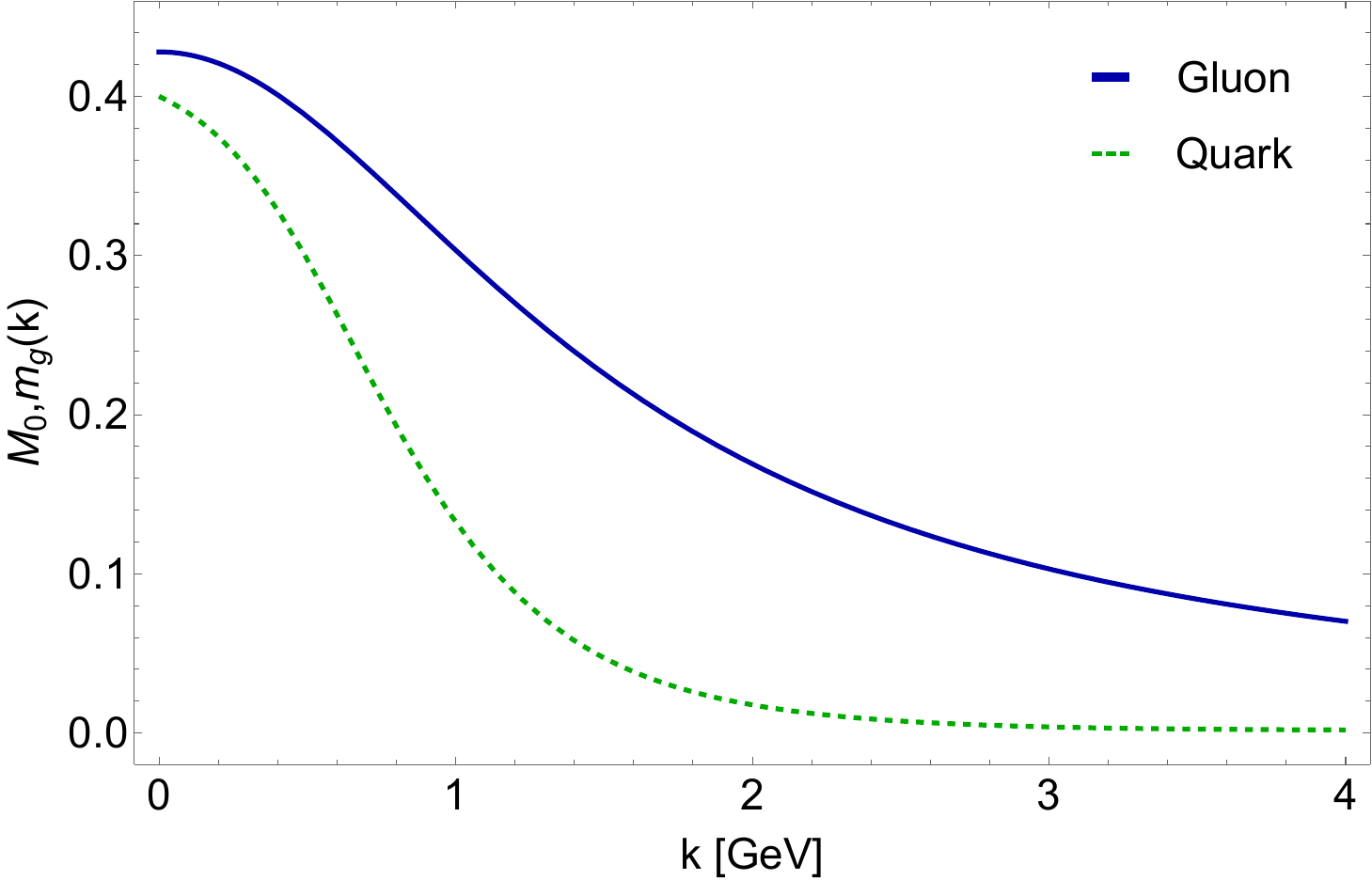} &
\includegraphics[width=0.6\textwidth]{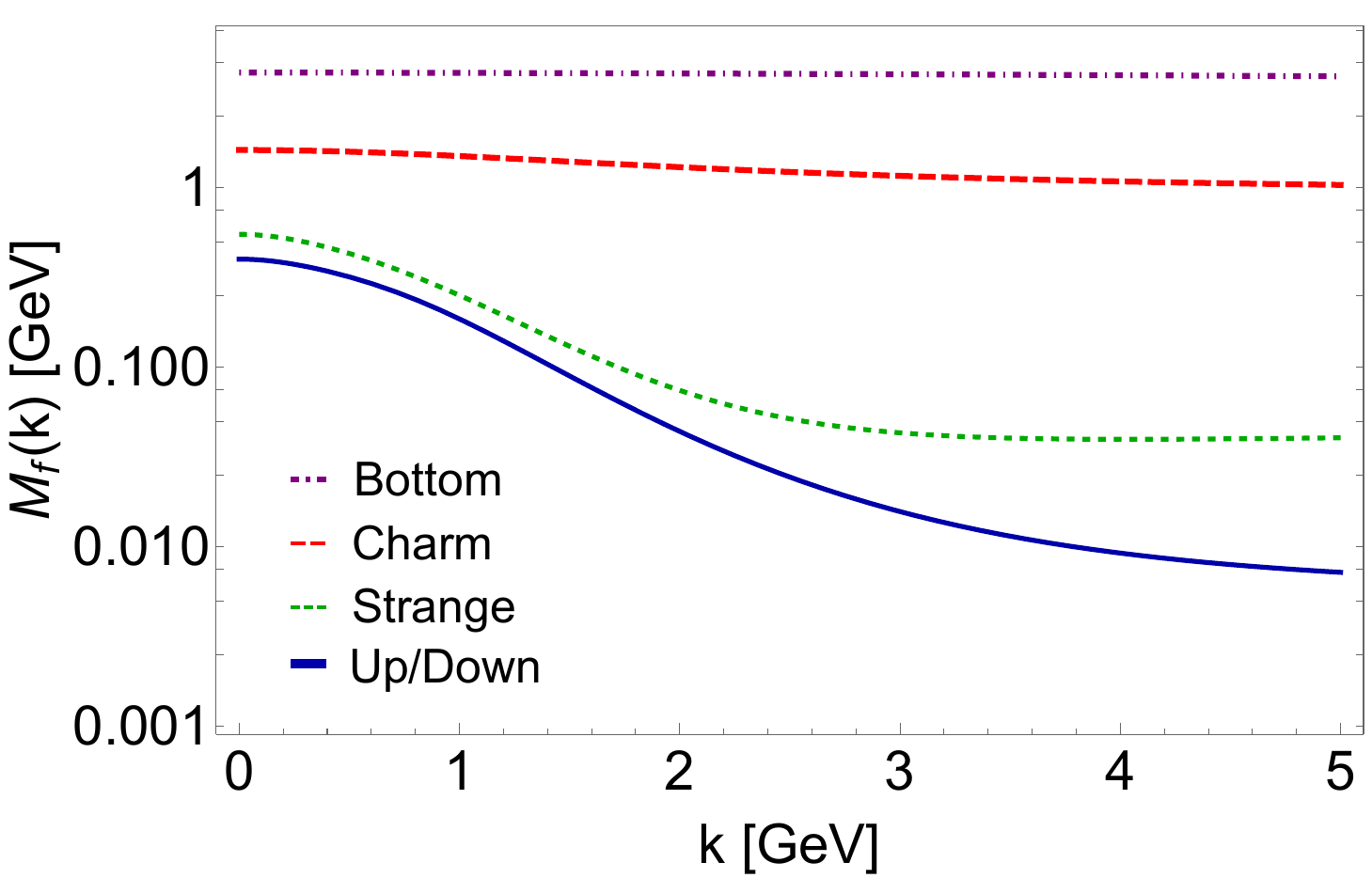}
\end{tabular}
\end{adjustwidth}
\caption{Running
 masses\,\cite{Raya:2024ejx}. (\textbf{Left}) Gluon effective running mass $m_g(k)$ (solid) and chiral-limit quark mass function $M_0(k)$ (dashed). (\textbf{Right}) Mass functions for different quark flavors. \label{fig:MassFuncs}}
\end{figure}

 Evidently, the bare case $S^{(0)}(p)$ is recovered when $Z(p^2),M(p^2)\to 1,m^{\text{bm}}$, with $m^{\text{bm}}$ the Lagrangian bare mass. All non-perturbative effects are thus encoded in the dressing functions $Z(p^2)$ and $M(p^2)$. The latter is commonly referred to as the \emph{mass function}
 Its profile provides a clear signature of mass generation in the matter sector, via DCSB\,\cite{Serna:2018dwk,Sultan:2018tet,Chang:2021vvx}. This is illustrated in Figure\,\ref{fig:MassFuncs}, which also shows a comparison with the gluon effective running mass\,\cite{Cui:2019dwv}---another manifestation of the non-Abelian character of QCD\,\cite{Ferreira:2025anh,Binosi:2022djx}.
 
The properties of mesons as quark–antiquark bound states can be derived through the Poincaré-covariant Bethe–Salpeter (BS) equation:
\vspace{-6pt}
\begin{equation}
    \label{eq:BSEdef}
    \Gamma_\textbf{X}(p;P)=\int_q \mathcal{K}^{(2)}(q,p;P)[S_q(q+\eta P)\Gamma_\textbf{X}(q;P)S_{\bar{h}}(q-(1-\eta)P);P)].
\end{equation}

Here, $\Gamma_\textbf{X}(p;P)$ denotes the $\textbf{X}$ meson BS amplitude, with the first entry ($p$) corresponding to the relative momentum between the valence quark and antiquark, and the second entry ($P$) the total momentum of the meson. For convenience, we have made explicit the valence quark ($q$) and antiquark ($\bar{h}$) flavors. The number of Dirac covariants that characterize this amplitude depends on the meson's quantum numbers\,\cite{Krassnigg:2010mh}. In particular, for the pseudoscalar case\,\cite{Maris:1997tm},
\begin{equation}
    \label{eq:BSEps}
    \Gamma_{\textbf{P}}(p;P)=\gamma_5[i E_{\textbf{P}}(p;P)+\gamma \cdot P\, F_{\textbf{P}}(p;P)+\gamma\cdot k\, k\cdot P\, G_{\textbf{P}}(p;P)+\sigma_{\mu\nu}k_\mu P_\nu\, H_{\textbf{P}}(p;P)].
\end{equation}

The quantity $\eta\in(0,1)$ appearing in Equation\,\eqref{eq:BSEdef} defines the momentum partitioning. In any Poincaré-covariant framework, physical observables are independent of it. In addition, $\mathcal{K}^{(2)}(q,p;P)$ denotes the 2-body kernel containing all possible interactions between the valence quark and antiquark. Being a homogeneous equation, \mbox{Equation\,\eqref{eq:BSEdef} adopts} the form of an eigenvalue equation and admits solutions only for discrete values $P_i$, satisfying $P_i^2 = -m_{\mathbf{P}i}^2$, where the lowest eigenvalue corresponds to the ground-state meson mass $m_{\mathbf{P}}$ and the higher ones to its radial excitations (e.g., Refs.\,\cite{Qin:2011xq,Mojica:2017tvh,Hilger:2017jti,Chang:2019eob}). Note we use the Euclidean convention. Furthermore, the resulting BS amplitudes must be canonically normalized in order to compute physical observables\,\cite{Nakanishi:1969ph}.

It is worth noticing that the 1- and 2-body kernels in the gap and BS equations are intrinsically linked and, as established before, embedded within an infinite tower of coupled equations. Any tractable problem then requires systematically decoupling them from the remainder of the tower. In the quark DSE, Equation\,\eqref{eq:quarkPropQCD}, this is typically done by employing a physically and mathematically motivated structure for the QGV, which often results in the gluon propagator becoming an effective one\,\cite{Binosi:2016rxz,Binosi:2016wcx,Serna:2018dwk,Sultan:2018tet,Chang:2021vvx}. Symmetry principles tightly relate $\mathcal{K}^{(1)}$ and $\mathcal{K}^{(2)}$, guaranteeing, for instance, the emergence of the pion as the Nambu–Goldstone boson in the chiral limit\,\cite{Munczek:1994zz,Bender:1996bb}. 


The way to determine $\mathcal{K}^{(1)}$ and $\mathcal{K}^{(2)}$ given an Ansatz for the QGV has been addressed in, e.g.,\,\cite{Chang:2009zb,Qin:2020jig}. For the sake of brevity, we limit ourselves to showing the so-called rainbow-ladder (RL) truncation, which is the leading-order symmetry-preserving truncation\,\cite{Munczek:1994zz,Bender:1996bb,Chang:2009zb,Qin:2020jig}. Consider the kernel:
\begin{equation}
    \label{eq:RL}
    \mathcal{K}^{(\text{RL})}(q,p):=-\frac{4}{3} Z_2D_{\mu\nu}^{\text{RL}}(p-q) \gamma_\mu\otimes  \gamma_\nu(p,q)\,;\,D_{\mu\nu}^{\text{RL}}(k):=\mathcal{G}(k)\left(\delta_{\mu\nu}-\frac{k_\mu k_\nu}{k^2}\right),
\end{equation}
with $\mathcal{G}(k)$ being an effective coupling, such as the well-known Maris–Tandy\,\cite{Maris:1999nt} and \mbox{Qin–Chang interaction} models\,\cite{Qin:2011dd}. Thus, the setting $-\mathcal{K}^{(1)}(q,p)=\mathcal{K}^{(2)}(q,p)=\mathcal{K}^{(\text{RL})}(q,p)$ defines the RL truncation. 

\textls[-15]{Additional truncation schemes have been developed to incorporate, among other features, the dynamically generated quark anomalous chromomagnetic \mbox{moment\,\cite{Xu:2022kng,Xu:2026sjy,Binosi:2016rxz,Binosi:2016wcx},} the effects of the non-Abelian anomaly and flavor mixing\,\cite{Ding:2018xwy,Bhagwat:2007ha}, as well as meson-cloud and isospin-breaking effects\,\cite{Miramontes:2021xgn,Miramontes:2022mex,Miramontes:2026ffk,Bashir:2026keh}. Likewise, several extensions of the RL approximation have been proposed to address systems with pronounced flavor asymmetry\,\cite{Chang:2019eob,Serna:2020txe,Qin:2019oar,Xu:2024rew,daSilveira:2022pte}.} 

The quark–photon vertex (QPV) is another central element in the computation of electromagnetic FFs. Once a truncation scheme has been specified, the QPV may be obtained from an inhomogeneous BS equation whose structure parallels that of the corresponding $q\bar{q}$ vector
bound-state equation, with an inhomogeneous driving term proportional to the tree-level structure $\gamma_\mu$\,\cite{Maris:1999bh}. In practice, however, by exploiting symmetry constraints, kinematic limits, and other well-established properties (see, e.g.,\,\cite{Bermudez:2017bpx,Albino:2018ncl,Albino:2021rvj}), one can ultimately express the vertex in terms of the quark propagator dressing \mbox{functions\,\cite{Chang:2013nia,Raya:2015gva,Raya:2016yuj,Raya:2019dnh,Ding:2018xwy}.} \mbox{A proper} implementation ensures the validity of either approach\,\cite{Miramontes:2021exi}. 

\subsection{Elastic Form Factors}
\label{sec:elastic}
Let us consider a pseudoscalar meson $\textbf{P}$ composed of a $q(\bar{h})$-flavored valence quark (antiquark). As discussed before, the matrix element associated with the elastic process $\gamma^* \textbf{P} \to \textbf{P}$, Equation\,\eqref{eq:MatElemEFF}, is described by a single form factor $F_\textbf{P}(Q^2)$. We shall consider the so-called impulse approximation, wherein the external photon couples directly to a single valence constituent while the other remains as a spectator. Thus, 
\begin{equation}
\label{eq:matelem}
\Lambda_\mu^{\gamma\textbf{P}\textbf{P}} = 2 K_\mu F_\textbf{P}(Q^2) \equiv K_\mu[e_q F_\textbf{P}^q(Q^2)+e_{\bar{h}} F_\textbf{P}^{\bar{h}}(Q^2)].
\end{equation}

The two terms $F_\textbf{P}^{q(\bar{h})}(Q^2)$ correspond to the valence-quark (antiquark) contribution, with electric charges $e_{q,\bar{h}}$ expressed in terms of the positron charge. Each of these valence components is derived from the corresponding triangle diagram, for instance\,\cite{Chang:2013nia}:
\begin{eqnarray}
    K_\mu F_{\textbf{P}}^q(Q^2) = \text{tr}\int_q\,\chi_{\mu}^\gamma(q+p_o,q+p_i) \Gamma_{\textbf{P}}(q+p_i/2;p_i)S_{\bar{h}}(q)\Gamma_{\textbf{P}}(q+p_o/2;-p_o);\label{eq:defEFF2}
\end{eqnarray}
here, $\chi_\mu^\gamma(r,s) := S(r)\Gamma_\mu^\gamma(r,s)S(s)$ is the unamputated QPV\,\cite{Raya:2015gva,Chang:2013nia}, and $\text{tr}$ indicates a trace over Dirac and color indices. The rest of the pieces carry the meanings explained in previous sections, and the antiquark component is obtained in a similar manner. Recall that $p_i(p_o)$ denotes the incoming (outgoing) meson momentum. The on-shell condition then entails $p_i^2=p_o^2=-m_\textbf{P}^2$.

\begin{figure}[H]%

\begin{adjustwidth}{-\extralength}{0cm}
\centering 
\begin{tabular}{lr}
\includegraphics[width=0.6\textwidth]{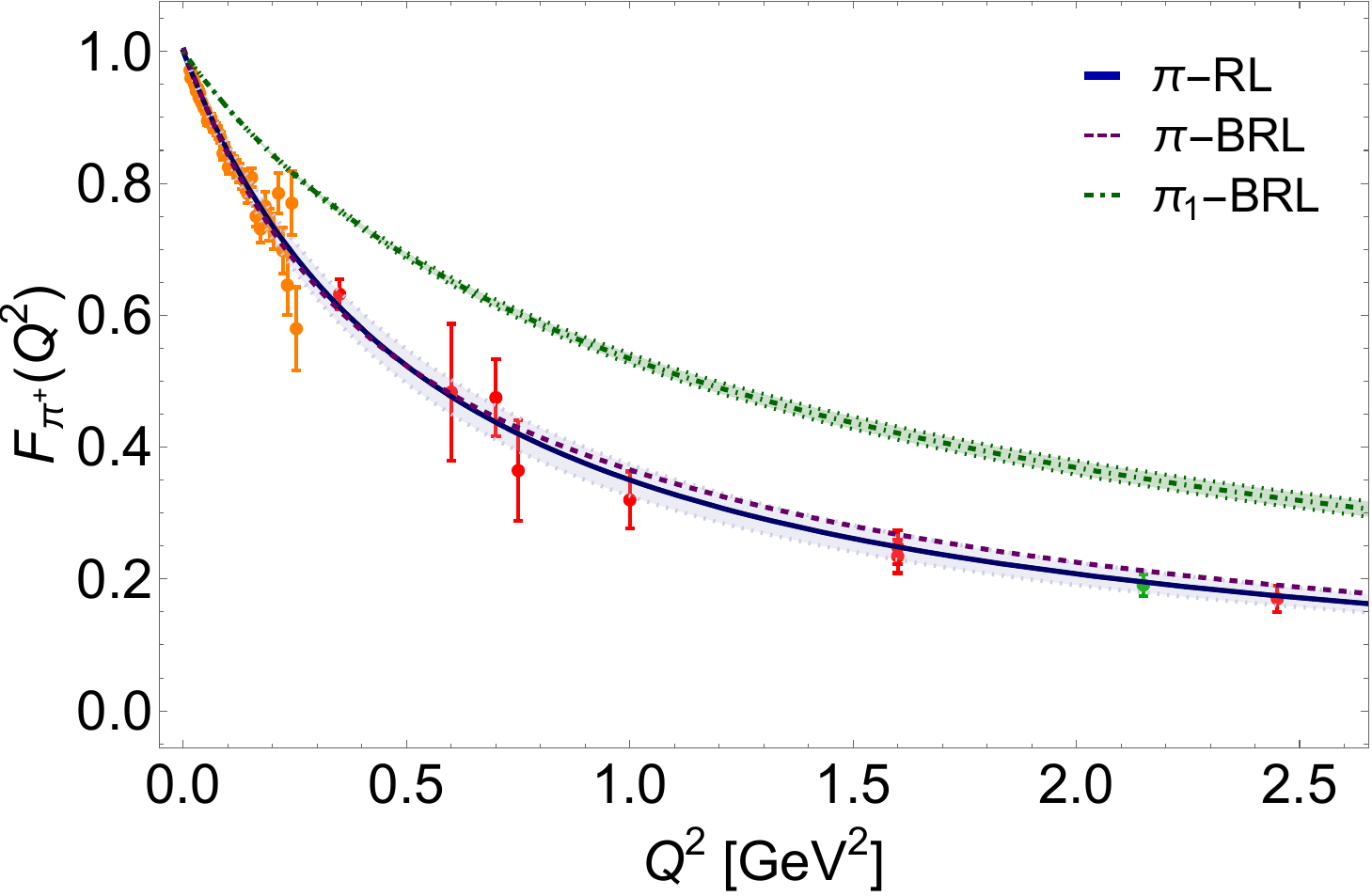} &
\includegraphics[width=0.6\textwidth]{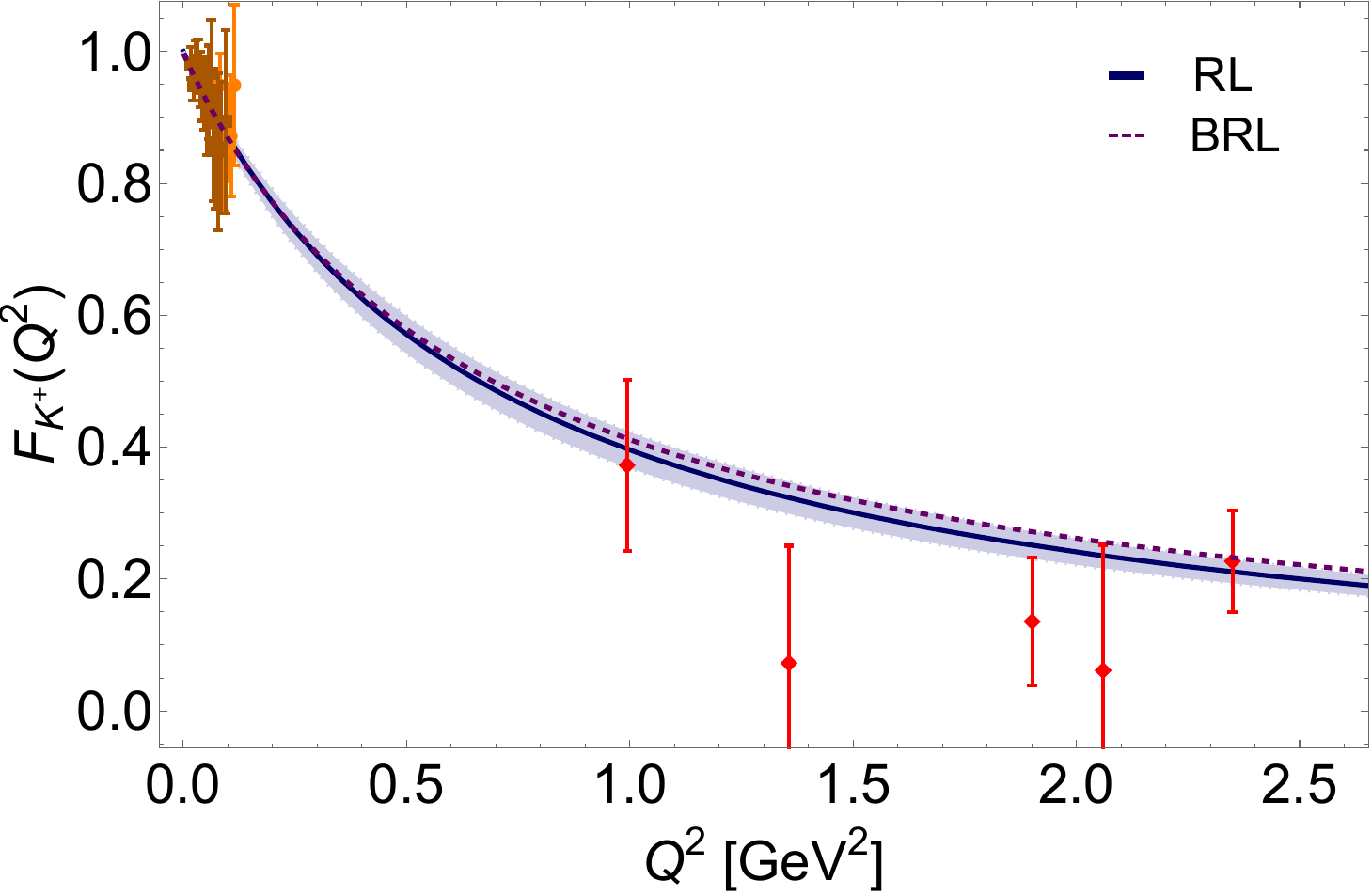}
\end{tabular}
\end{adjustwidth}
\caption{$\pi^+-K^+$
 elastic electromagnetic FFs. (\textbf{Left}) Ground-state pion ($\pi$) and its first radial excitation ($\pi_1$) calculated in the RL truncation and in the RL + meson-cloud scheme (BRL) from Refs.\,\cite{Miramontes:2021exi,Miramontes:2024fgo,Raya:2022ued}. (\textbf{Right}) Corresponding results for the ground-state $K^+$ meson. Pion data from \mbox{Refs.\,\cite{NA7:1986vav, Horn:2007ug, JeffersonLab:2008jve}} and kaon data from Refs.\,\cite{Dally:1980dj, Amendolia:1986ui, Carmignotto:2018uqj}.\label{fig:EFFlow}}
\end{figure}

The low-$Q^2$ behavior of the ground-state $\pi^+$ and $K^+$ elastic FFs, together with that of the first radial excitation of the pion, is illustrated in Figure\,\ref{fig:EFFlow}. Clearly, the theoretical predictions are in excellent agreement with the available experimental data, although the  coverage remains more limited for the $K^+$. The RL truncation results and those that incorporate meson-cloud effects (dubbed BRL) converge to essentially the same profiles\,\cite{Miramontes:2021exi}. This indicates that meson-cloud contributions are strongly suppressed in the spacelike-momentum regime, becoming increasingly negligible as $Q^2$ grows. Moreover, \mbox{Figure\,\ref{fig:EFFlow} highlights} the reliability of approaches in which the form factors are evaluated using a perturbation-theory integral representation (PTIR) for the relevant propagators and vertices\,\cite{Miramontes:2021exi,Gao:2017mmp,Chang:2013nia,Raya:2015gva}. {\mbox{A brief} explanation of the PTIR is provided in Appendix\,\ref{app:appendix}.}

It is worth emphasizing that the DSE framework is able to capture the radially excited nature of the $\pi_1:=\pi(1300)$ state\,\cite{Miramontes:2024fgo,Li:2016dzv}. First, the solution of the corresponding BS equation naturally identifies it as the second eigenvalue, thereby generating the characteristic node in the wave function. Moreover, as a corollary of correctly reproducing the Nambu–Goldstone boson nature of the ground-state pion, the framework automatically yields a decay constant $f_{\pi_1}$ that vanishes identically in the chiral limit and remains small for physical values of current quark mass. This characteristic behavior provides \mbox{a stringent} consistency check on any theoretical framework intended to describe radially excited pseudoscalar mesons\,\cite{Dominguez:1976ut,Dominguez:1977nt}.

The corresponding charge radii, Equation\,\eqref{eq:ChargeR}, are (in fm)
\begin{equation}
    r_{\pi^+}=0.66(1)\,,\,r_{K^+}=0.59(1)\,,\,r_{\pi_1}=0.58(1).
\end{equation}

These values not only agree with available phenomenological determinations\,\cite{Cui:2022fyr,ParticleDataGroup:2024cfk} but also encode other interesting aspects. For instance, the charged kaon-to-pion ratio is on the order of $r_{K^+}/r_\pi^+\approx0.89\approx f_\pi/f_K$, consistent with recent analyses\,\cite{Raya:2024ejx,Raya:2024glv}, and traceable to the slightly larger contribution of the Higgs mechanism to the kaon mass budget. On the other hand, despite being nearly an order of magnitude heavier, the radially excited pion has an electromagnetic size comparable to that of the ground state, with $r_{\pi_1}/r_\pi^+\approx0.86$. {This result is somewhat counterintuitive, as one might naively anticipate a larger charge radius for a radially excited state. Its origin may, however, be sensitive to the details of the underlying bound-state description of the excited state, including the model assumptions and parameters\,\cite{Holl:2005vu}. A more robust truncation scheme would reduce this dependence.
}

\begin{figure}[H]%

\begin{adjustwidth}{-\extralength}{0cm}
\centering 
\begin{tabular}{lr}
\includegraphics[width=0.6\textwidth]{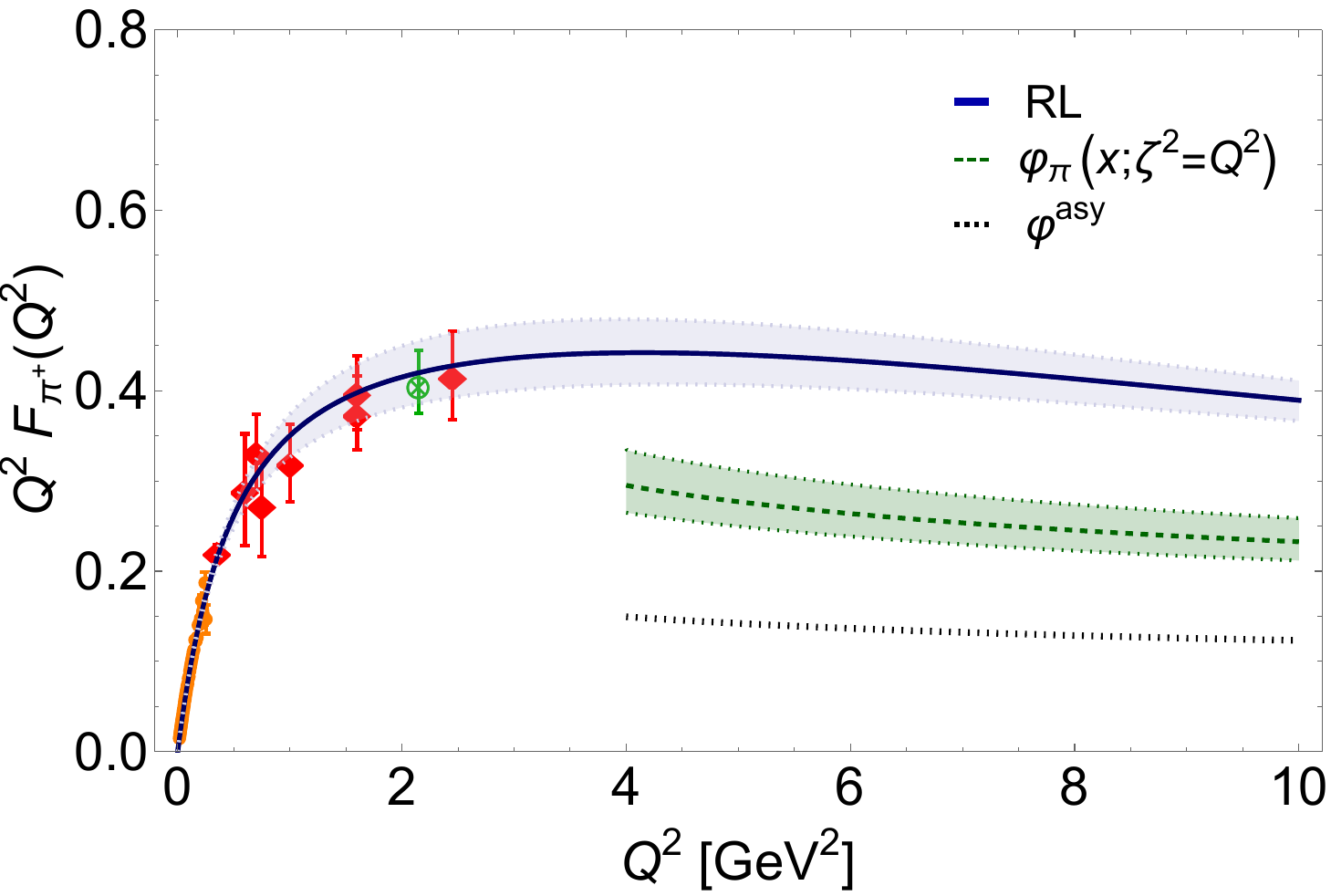} &
\includegraphics[width=0.6\textwidth]{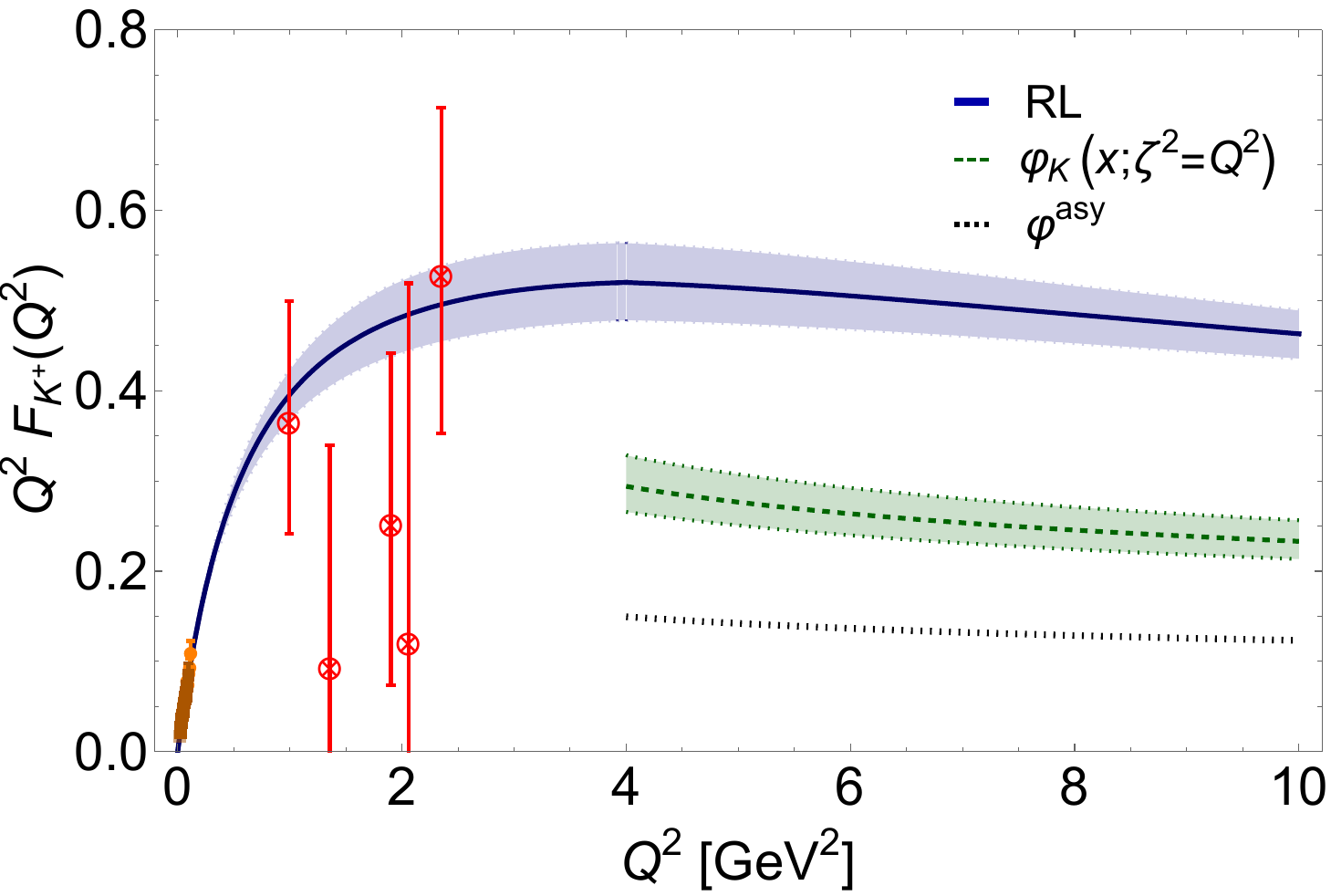}
\end{tabular}
\end{adjustwidth}
\caption{$\pi^+-K^+$
 elastic electromagnetic FFs. (\textbf{Left}) Pion RL result, compared with the hard-scattering formula expectations, Equation\,\eqref{eq:FPUV}. (\textbf{Right}) Corresponding results for the kaon. The DAs entering the hard-scattering formulae are taken from Ref.\,\cite{Raya:2024ejx}. Pion data from Refs.\,\cite{NA7:1986vav, Horn:2007ug, JeffersonLab:2008jve} and kaon data from Refs.\,\cite{Dally:1980dj, Amendolia:1986ui, Carmignotto:2018uqj}.\label{fig:EFFlarge}}
\end{figure}

The large-$Q^2$ behavior of the $\pi^+-K^+$ elastic FFs is displayed in Figure\,\ref{fig:EFFlarge}. Since the DSE formalism is rooted in the QCD Lagrangian, it naturally incorporates the perturbative dynamics, and consequently, the scaling violations predicted by Equation\,\eqref{eq:FPUV} emerge explicitly in this regime. Access to this asymptotic domain, however, was first achieved through the PTIR framework introduced in\,\cite{Chang:2013nia}, which enables reliable evaluations of the FFs at arbitrarily large spacelike momenta. This capability was subsequently exploited in studies of the kaon elastic FFs\,\cite{Gao:2017mmp} and of the complete set of ground-state pseudoscalar two-photon transition FFs\,\cite{Raya:2015gva,Raya:2016yuj,Raya:2019dnh,Ding:2018xwy}. These results are discussed next.

\subsection{Two-Photon Transitions}
Adopting the impulse approximation, we may express the two-photon transition form factor introduced in Equation\,\eqref{eq:MatElemTFF} as follows\,\cite{Raya:2015gva}:
\begin{eqnarray}
    \mathcal{T}_{\mu\nu}(Q_1,Q_2)
    =e^2\mathcal{Q}^2_{\textbf{P}}\mbox{tr} \int_{q} i\chi_{\mu}^\gamma(q,q+Q_1)\Gamma_{\textbf{P}}(q+Q_1,q-Q_2)  S(q-Q_2)i \Gamma_{\nu}^q(q-Q_2,q);\label{eq:Tmunu}
\end{eqnarray}
recall $Q_{1,2}$ are the photon momenta, such that $(Q_1+Q_2)^2=-m_\textbf{P}^2$ for an on-shell meson; the factor $\mathcal{Q}_\textbf{P}^2$, introduced in connection with Equation\,\eqref{eq:GPUV}, accounts for the meson's flavor structure. The $\eta-\eta'$ case requires minor modifications\,\cite{Ding:2018xwy}.

\begin{figure}[H]%

\begin{adjustwidth}{-\extralength}{0cm}
\centering 
\begin{tabular}{lr}
\includegraphics[width=0.6\textwidth]{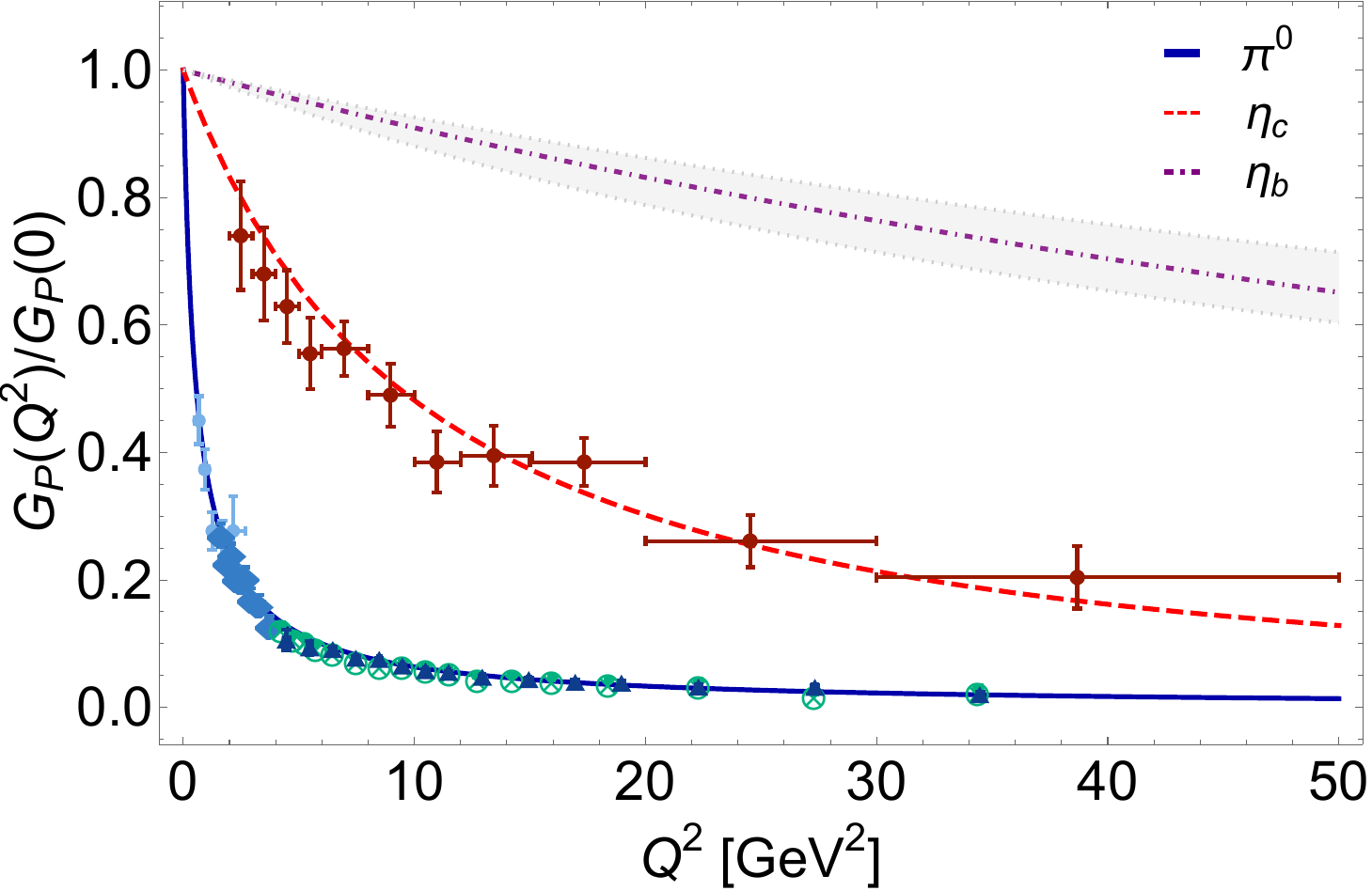} &
\includegraphics[width=0.6\textwidth]{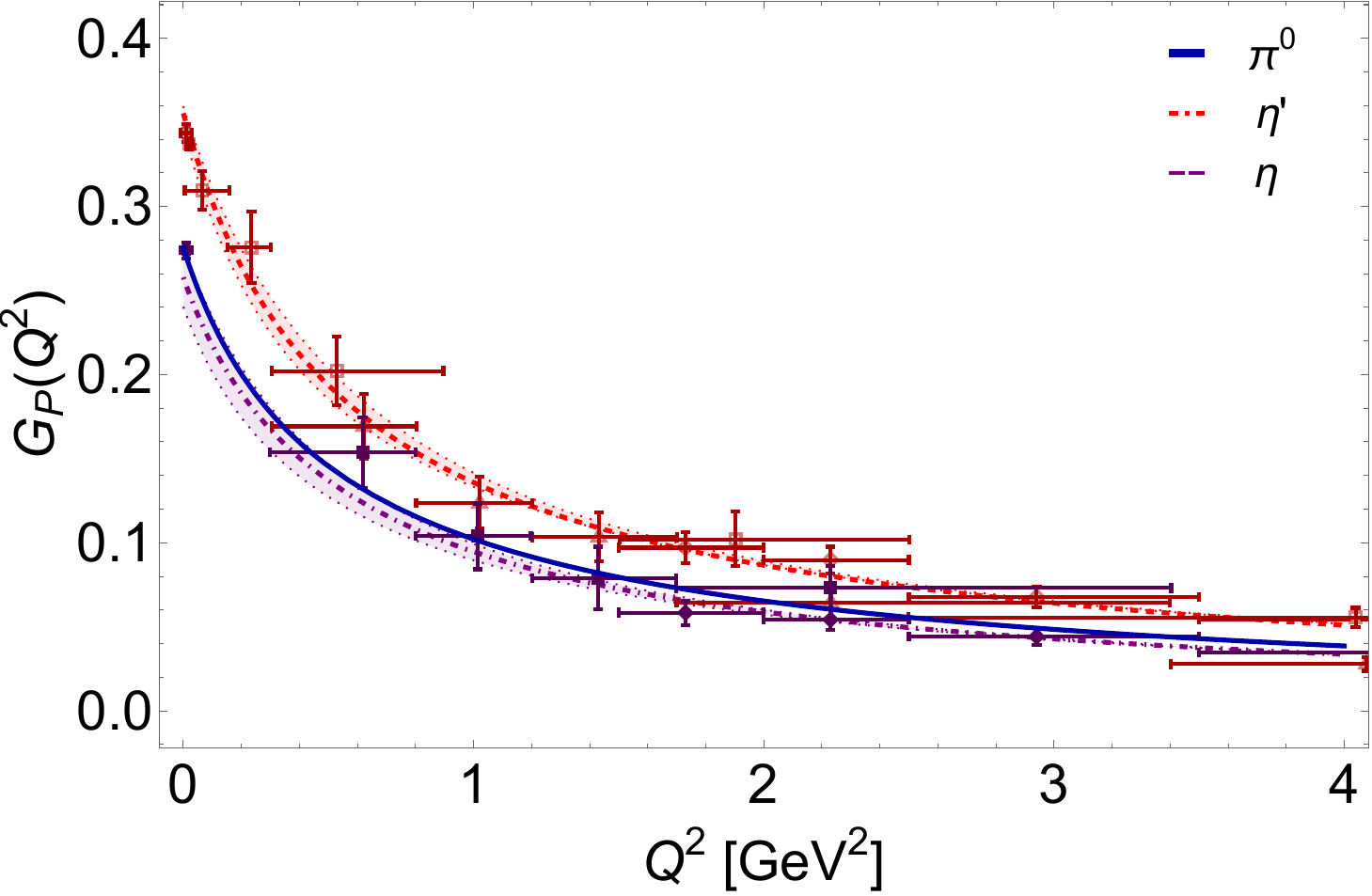}
\end{tabular}
\end{adjustwidth}
\caption{$\gamma^*\gamma$
 transition FFs of ground-state pseudoscalar mesons\,\cite{Raya:2015gva,Raya:2016yuj,Raya:2019dnh,Ding:2018xwy}. (\textbf{Left}) Heavy quarkonia $\eta_c$ and $\eta_b$. (\textbf{Right}) The intertwined $\eta-\eta'$ system. In both panels, the $\pi^0$ prediction is shown for comparison. Experimental data: $\pi^0$\,\cite{CELLO:1990klc,CLEO:1997fho,BaBar:2009rrj,Belle:2012wwz}, $\eta-\eta'$\,\cite{CELLO:1990klc,CLEO:1997fho,L3:1997ocz,BaBar:2011nrp}, and $\eta_c$\,\cite{BaBar:2010siw}. For the $\eta-\eta'$ case, we have also included the PDG values listed in Equation\,\eqref{eq:ValuesG0}. Since no data are currently available for $\eta_b$, we have included the nonrelativistic QCD expectation (gray band) as a benchmark\,\cite{Feng:2015uha}. \label{fig:TFFlow}}
\end{figure}

The two-photon transition FFs for all ground-state neutral pseudoscalar mesons are displayed in Figure\,\ref{fig:TFFlow}. Focusing on the left panel, displaying the $\pi^0,\eta_c,\eta_b$ transition FFs, we observe a clear trend: As the system becomes heavier, the form factors exhibit \mbox{a progressively} slower falloff. This behavior reflects the changing balance between EHM and Higgs-generated quark masses in shaping the hadron's structure. As the latter mechanism becomes increasingly dominant, the smaller slope of the FF indicates the meson becomes more compact\,\cite{Raya:2024ejx,Raya:2024glv}.

The right panel of Figure\,\ref{fig:TFFlow} shows transition FFs of the three light mesons, namely $\pi^0,\eta,\eta'$. The close similarity between the $\pi^0$ and $\eta$ transition FFs, in both shape and magnitude, is a natural consequence of the fact that both mesons can be regarded as (pseudo) Nambu–Goldstone bosons of DCSB. The $\eta'$ meson, on its part, lies outside the Nambu–Goldstone multiplet owing to its intimate connection with the non-Abelian anomaly; consequently, its transition FF features a larger magnitude and a somewhat slower falloff. The following values are found at the soft limit (in GeV$^{-1}$):
\begin{eqnarray}
\label{eq:ValuesG0DSE}
    &&\bar{G}^0_\pi=0.2753(31)\,,\,\bar{G}^0_{\eta}=0.2562(170)\,,\bar{G}^0_{\eta'}=0.3495(60),\\
    &&\bar{G}^0_{\eta_c}=0.0705(40)\,\bar{G}^0_{\eta_b}=0.0038(2)
    ,
\end{eqnarray}
whereas the slope parameters read (in GeV$^{-2}$):
\begin{equation}
\label{eq:slopeDSE}
    b_{\pi^0}=1.74(2)\,,\,b_\eta=1.86(11)\,,\,b_{\eta'}=1.66(8)\,,b_{\eta_c}=0.31(1)\,,b_{\eta_b}=0.10(1)
    .
\end{equation}

All these values are fully consistent with the phenomenological results presented in the previous section. {Moreover, the recent DSE analysis from Ref.\,\cite{Ye:2026zkw} corroborates these patterns, within a numerical framework that enables the comparison between a conventional rest-frame calculation and a moving-frame one, complemented by a Schlessinger point method extrapolation (e.g., Refs.\,\cite{Schlessinger:1966zz,Cui:2022fyr}). The latter is required to access sufficiently large photon virtualities. As discussed in Ref.\,\cite{Xu:2024rew} in the context of strong decays, moving-frame calculations offer a promising alternative to mitigate the uncertainties associated with the interpolation and extrapolation of numerically obtained BS amplitudes. Controlling these uncertainties is particularly important for precision studies of the muon $g-2$.}

\begin{figure}[H]%

\begin{adjustwidth}{-\extralength}{0cm}
\centering 
\begin{tabular}{lr}
\includegraphics[width=0.6\textwidth]{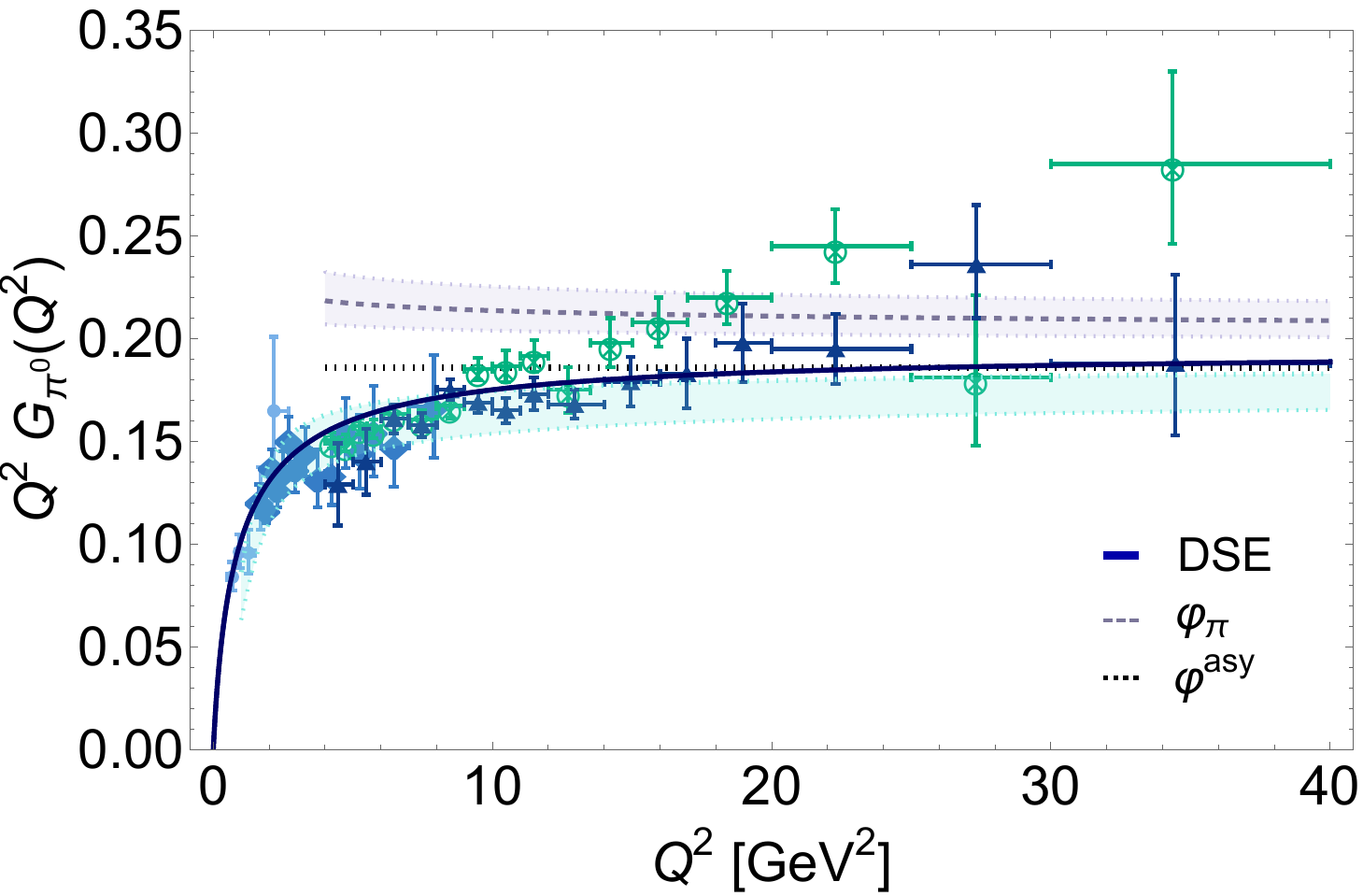} &
\includegraphics[width=0.6\textwidth]{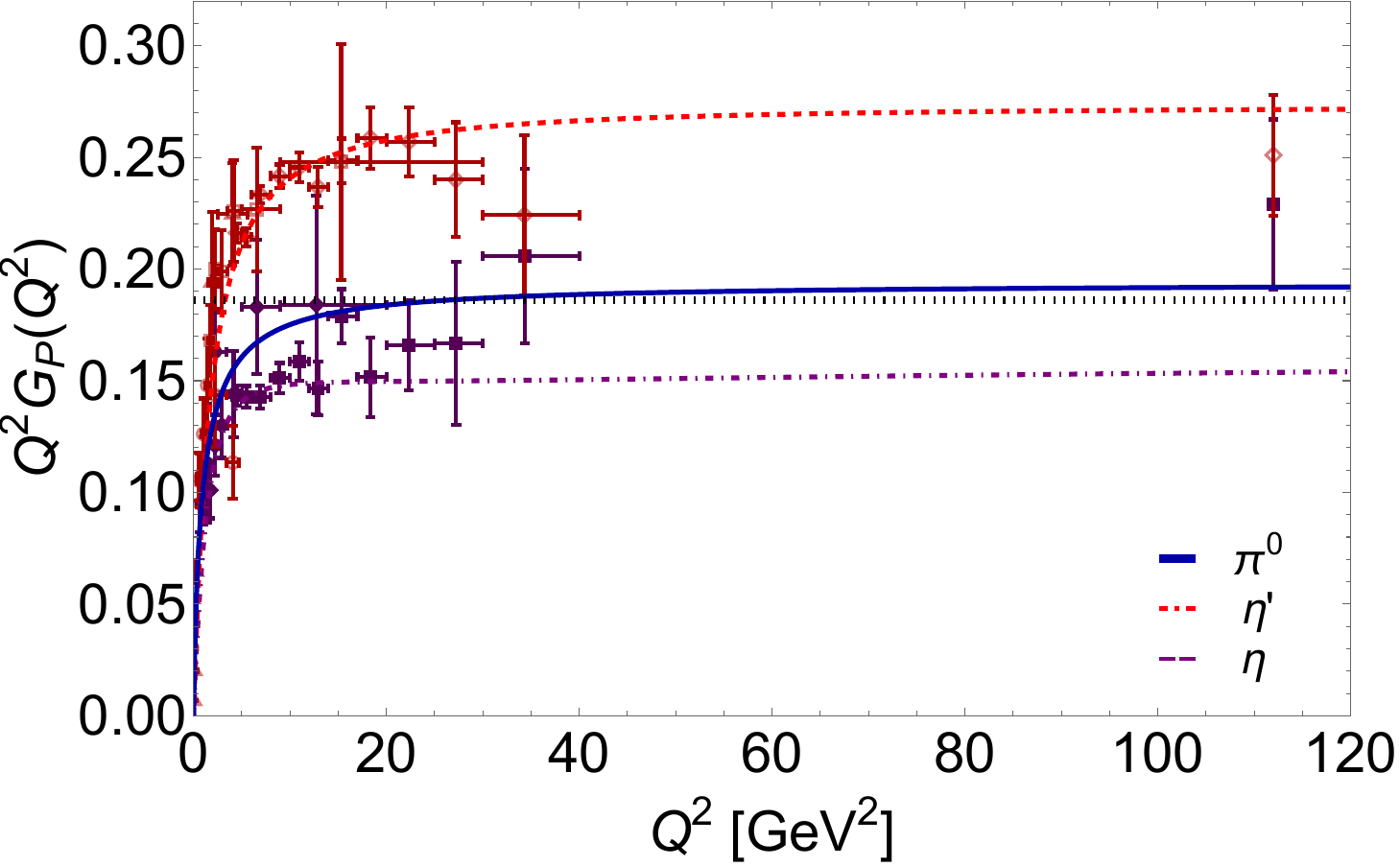}
\end{tabular}
\end{adjustwidth}
\caption{$\gamma^*\gamma$
 transition FFs of light pseudoscalar mesons\,\cite{Raya:2015gva,Raya:2016yuj,Raya:2019dnh,Ding:2018xwy}. (\textbf{Left}) Neutral pion transition FF. The solid curve is the DSE prediction in the impulse approximation. The dashed gray curve and its uncertainty band are obtained from the hard-scattering formula, Equation~\eqref{eq:GPUV}, using the pion DA from Ref.\,\cite{Raya:2024ejx}; the dotted line shows the asymptotic limit $2f_\pi$, corresponding to the asymptotic DA, Equation~\eqref{eq:asymDA}. The light-blue band is the analysis of Ref.~\cite{Bakulev:2012nh}. (\textbf{Right}) The intertwined  $\eta$ (purple, dot-dashed) and $\eta'$ (red, dashed) system. The pion result and its asymptotic limit have been included for comparison. Experimental data from Refs.\,\cite{CELLO:1990klc,CLEO:1997fho,BaBar:2009rrj,Belle:2012wwz,L3:1997ocz,BaBar:2011nrp,BaBar:2006ash}. \label{fig:TFFlarge}}
\end{figure}

Figure\,\ref{fig:TFFlarge} displays the scaled transition FFs of the light pseudoscalar mesons, $Q^2G_\textbf{P}(Q^2)$. The pion result, shown in the left panel, reveals a discrepancy that is not apparent from Figure\,\ref{fig:TFFlow}: The BaBar\,\cite{BaBar:2009rrj} and Belle\,\cite{Belle:2012wwz} measurements become increasingly incompatible as the photon virtuality grows. On its part, the DSE prediction is consistent with the Belle data, the asymptotic expectations from Equation\,\eqref{eq:GPUV}, and the light-cone sum rules analysis from\,\cite{Bakulev:2012nh}.

The right panel of Figure\,\ref{fig:TFFlarge} displays the corresponding results for the $\eta-\eta'$ mesons. Excellent agreement is found with all available experimental data, including the extraction at $Q^2=112\,\text{GeV}^2$\,\cite{BaBar:2006ash}, as well as with the corresponding asymptotic limits. The $\eta_c$ and $\eta_b$ results at large $Q^2$ have been omitted, as their considerably larger masses imply that the asymptotic regime is attained only at much higher virtualities. Further details can be found in Refs.\,\cite{Raya:2015gva,Ding:2018xwy,Raya:2016yuj,Raya:2019dnh}.

\vspace{-7pt}

A notable outcome of the present approach is that the asymptotic expectations are faithfully reproduced, namely:
\begin{eqnarray}
\label{eq:AsymDSE1}
    &&\bar{G}_\pi^{asy}=2f_\pi\,,\,\bar{G}_{\eta_c}^{asy}=(8/3)f_{\eta_c}\,,\bar{G}_{\eta_c}^{asy}=(2/3)f_{\eta_b},
\end{eqnarray}
where $f_\pi=0.092$, $f_{\eta_c}=0.26$, and $f_{\eta_b}=0.54$ (in GeV). For the $\eta-\eta'$ mesons, one finds
\begin{eqnarray}
\label{eq:AsymDSE2}
    \bar{G}_\eta^{asy}=0.158(13)\,,\,\bar{G}_{\eta'}^{asy}=0.285(11)
    ,
\end{eqnarray}
in line with the data-driven expectations quoted in Equation\,\eqref{eq:asydisp}. The asymptotic treatment of the $\eta-\eta'$ system requires special care due to the non-Abelian axial anomaly. In addition to the evolution of the meson DAs, one must account for the evolution of the gluon DAs and the scale dependence of the decay constants\,\cite{Agaev:2014wna}. The systematic way to achieve it within the DSE formalism is detailed in Ref.\,\cite{Ding:2018xwy}.

Thus, all transition form factors are consistent with both their low-energy behavior and the expected asymptotic limits, while exhibiting excellent agreement with the available experimental data. The only visible tension is with the large-$Q^2$ measurements reported by BaBar for the neutral pion transition FF\,\cite{BaBar:2009rrj}, which show an increasing trend and largely exceed the asymptotic expectations. It is also worth recalling that the same framework yields the electromagnetic form factors of the $\pi^+$ and $K^+$ in agreement with both experiment and the expected high-energy behavior.

With this extensive assortment of form factors at hand, we are now in a position to estimate the corresponding meson-pole and box contributions.

\section{HLbL Contributions Within DSEs}
\label{sec:HLbLDSE}

\subsection{Pseudoscalar-Pole and Box Contributions}
The transition FFs discussed in the previous section constitute the central hadronic input for the dominant pseudoscalar-pole contribution to HLbL scattering. Combined with the electromagnetic FFs discussed in Section\,\ref{sec:elastic}, they enable a quantitative determination of the leading hadronic contributions to $a_\mu$.

The box contribution is readily evaluated from Equation\,\eqref{eq:PiBox} once the required electromagnetic FFs are available. The pseudoscalar-pole contribution, on the other hand, is obtained from Equation\,\eqref{eq:amuPSpole} and necessitates a reliable parametrization of the numerically determined transition FFs. Ref.\,\cite{Masjuan:2017tvw} showed that the transition FFs of the light pseudoscalar mesons ($\pi^0$, $\eta$, and $\eta'$) can be accurately described by means of Canterbury approximants. Building upon this idea, Ref.\,\cite{Raya:2019dnh} employed the parametrization:
\begin{equation}
    G_\textbf{P}^{light}(x,y)=\frac{\bar{G}^0_\textbf{P}+a_{10}(x+y)+a_{01}xy}{1+b_{10}(x+y)+b_{01}xy+b_{11}(x+y)xy}.
\end{equation}

In addition to allowing for an accurate characterization of the numerical results, this representation permits us to reproduce the correct asymptotic behavior in both the single- and double-virtual limits, provided that
\begin{equation}
    a_{01}=(2/3)b_{11} \bar{G}_{\textbf{P}}^{asy}\,,\,a_{10}=b_{20}(1+\delta_{BL})\bar{G}_{\textbf{P}}^{asy},
\end{equation}
where $\bar{G}_{\textbf{P}}^{\rm asy}$ denotes the asymptotic limit defined in Equation\,\eqref{eq:asymTFF}. The exact asymptotic behavior is recovered for $\delta_{\rm BL}=0$, while allowing $\delta_{\rm BL}\neq0$ provides additional flexibility to improve the quality of the interpolation without altering the large-$Q^2$ power-law behavior.

For the case of $\eta_c$ and $\eta_b$, whose massiveness pushes away the asymptotic domain, the following characterization is sufficient:
\begin{equation}
    G_\textbf{P}^{heavy}(x,y)=\frac{\bar{G}^0_\textbf{P}M_{V_0}^4+c_{10}(x+y)}{(x+M_{V_0}^2)(y+M_{V_0}^2)}.
\end{equation}

Additional details and all parameters required to reproduce the FFs described herein can be found in Ref.\,\cite{Raya:2019dnh}.
The pseudoscalar-pole contributions obtained within the present DSE approach, along with a recapitulation of relevant low-energy quantities and asymptotic limits, are collected in Table\,\ref{tab:amuNeutral}. Corresponding results for the pseudoscalar-box contributions are shown in Table\,\ref{tab:amuCharged}. Taking these results together, one finds
\begin{equation}
\label{eq:amuPme}
a_\mu^{\textbf{P}-\mathrm{pole}}=90.6(3.9)\times10^{-11},\qquad
a_\mu^{\textbf{P}-\mathrm{box}}=-16.1(0.2)\times10^{-11}.
\end{equation}

For comparison, we recall the currently accepted values quoted in Equation\,\eqref{eq:currAmu} and examined deeply in the 2025 White Paper\,\cite{Aliberti:2025beg}:
\begin{equation}
    a_\mu^{\pi^0,\eta,\eta'-\text{poles}}=91.2_{-2.4}^{+2.9}\times10^
    {-11}\,,\,a_\mu^{\pi,K-\text{box}}=-16.4(0.2)\times10^{-11}.
\end{equation}

Evidently, our findings are fully consistent with this consensus. {A breakdown of the individual contributions is provided in Tables\,\ref{tab:amuNeutral} and\,\ref{tab:amuCharged}. The box contribution had previously been studied within the DSE framework in Ref.\,\cite{Eichmann:2019bqf}, yielding an identical result. 

For the pseudoscalar-pole contribution, it is worth distinguishing between the light- and heavy-meson sectors. We obtain
\begin{equation}
\label{eq:amuDiss}
    a_\mu^{\pi^0,\eta,\eta'-\mathrm{poles}}=89.7(4.8)\times10^{-11}\,,\,a_\mu^{\eta_c,\eta_b-\mathrm{poles}}\approx a_\mu^{\eta_c-\mathrm{pole}}=0.9(0.1)\times 10^{-11}.
\end{equation}

Other DSE results for the light-meson sector, both earlier\,\cite{Eichmann:2019tjk} and more recent\,\cite{Ye:2026zkw}, fall within the same range, respectively:
\begin{equation}
    a_\mu^{\pi^0,\eta,\eta'}=91.6(1.9)\times 10^{-11}, \,a_\mu^{\pi^0,\eta,\eta'}=84.3(0.8)\times 10^{-11}\,.
\end{equation}

In Ref.\,\cite{Eichmann:2019tjk}, the $\eta-\eta'$ system is treated self-consistently within the RL truncation, with a mixing scheme subsequently adopted. In contrast, as in our work (i.e.,\,Ref.\,\cite{Raya:2019dnh}), the recent exploration in Ref.\,\cite{Ye:2026zkw} incorporates the effects of the non-Abelian anomaly at the interaction kernel level. As discussed above, the latter employs a moving-frame formulation to reduce the numerical uncertainties associated with the evaluation of BS amplitudes, complemented by a Schlessinger point method extrapolation to access sufficiently large photon virtualities.}

It is worth noting that the DSE framework has, for the first time, enabled the contributions from the heavy-quarkonia sector to be computed. This analysis revealed that the $\eta_c$ contribution is comparable in magnitude to the current uncertainty associated with the light-pseudoscalar contribution, whereas the corresponding contribution from the $\eta_b$ is entirely negligible{. See Equation\,\eqref{eq:amuDiss}. These findings were corroborated in the recent analysis from\,\cite{Ye:2026zkw}, which, to the best of our knowledge, is the only other calculation to include heavy-quarkonia pole contributions.}

For the first time too, the contribution stemming from the first radially excited pion state, $\pi_1$, was also evaluated\,\cite{Miramontes:2021exi}. Interestingly, although the radially excited pion exhibits features that markedly distinguish it from its ground-state counterpart, including a node in its wave function and a decay constant of only a few MeV, the elastic FF of the $\pi_1$ has essentially the same shape as that of the ground-state pion. Consequently, its contribution to $a_\mu$ is governed predominantly by the scaling relation $a_\mu^{\pi_1-\text{pole}} \sim a_\mu^{\pi-\text{pole}}(m_\pi/m_{\pi_1})^3$, becoming negligible.

The results presented in Tables\,\ref{tab:amuNeutral} and\,\ref{tab:amuCharged}, regarding the pseudoscalar-pole and box HLbL components, are taken from Refs.\,\cite{Raya:2019dnh,Raya:2022ued,Miramontes:2021exi}. To date, these constitute the first and only comprehensive evaluation in which all ground-state neutral pseudoscalar mesons and the charged $\pi^+$ and $K^+$, as well as the first excited state of the pion, are treated within a single, unified framework. Remarkably, the same framework has also been employed to study quantities that may appear unrelated to $a_\mu$, yet are intimately connected to the underlying structure of hadrons. These include, among others, parton distribution functions and generalized parton distributions\,\cite{Raya:2021zrz,Raya:2026gwg,Cui:2020tdf}. {It is worth emphasizing that our assessment of the pseudoscalar-pole and box contributions\,\cite{Raya:2019dnh,Miramontes:2021exi,Miramontes:2024fgo,Raya:2022ued} and the independent DSE determinations from Refs.\,\cite{Eichmann:2019bqf,Eichmann:2019tjk,Ye:2026zkw}, despite relying on different implementations, share a common foundation in QCD. Their mutual consistency is therefore not accidental, but a reflection of the robustness and predictive power of the DSE approach}.

\begin{table}[H] 
\small
\caption{Pseudoscalar-pole
 contributions to $a_\mu$ within the DSE approach\,\cite{Raya:2019dnh,Raya:2022ued}. For completeness, the low-energy quantities associated with the corresponding transition FFs, $b_\textbf{P}$ and $\bar{G}_{\textbf{P}}^0$, together with their asymptotic limits, $\bar{G}_{\textbf{P}}^{asy}$, are also listed. In the present DSE study, $f_\pi=0.092$, $f_{\eta_c}=0.26$, and $f_{\eta_b}=0.54$ GeV.
\label{tab:amuNeutral}}
\begin{tabularx}{\textwidth}{CCCCC}
\toprule
	& \boldmath{$b_\textbf{P}/\text{\textbf{GeV}}^{-2}$} & \boldmath{$\bar{G}^0_\textbf{P}/\text{\textbf{GeV}}^{-1}$}	& \boldmath{$\bar{G}_\textbf{P}^{asy}/\text{\textbf{GeV}}$} & \boldmath{$a_\mu^{\textbf{P}-pole} \times 10^{-11}$} \\
\midrule
$\pi^0$ & $1.74(2)$ & $0.2753(31)$ & $2f_\pi$ & $61.4(2.1)$  \\
$\eta$ & $1.86(11)$ & $0.2562(170)$ & $0.158(13)$ & $14.7(1.0)$  \\
$\eta'$ & $1.73(4)$ & $0.3495(60)$ & $0.285(11)$ & $13.6(0.8)$  \\
$\eta_c$ & $0.31(1)$ & $0.0705(40)$ & $(8/3)f_{\eta_c}$ & $0.9(0.01)$  \\
$\eta_b$ & $0.10(1)$ & $0.0038(02)$ & $(2/3)f_{\eta_b}$ & $0.0026(0.0001)$  \\
\bottomrule
\end{tabularx}
\end{table}
\vspace{-9pt}

\begin{table}[H] 
\small
\caption{Pseudoscalar-box
 contributions to $a_\mu$ within the DSE approach\,\cite{Raya:2022ued,Miramontes:2021exi,Miramontes:2024fgo}. For completeness, the corresponding charge radii $r_\textbf{P}$ are also shown.
\label{tab:amuCharged}}
\begin{tabularx}{\textwidth}{CCC}
\toprule
	& \boldmath{$r_\textbf{P}\textbf{/}\text{\textbf{fm}}$} & \boldmath{$a_\mu^{\textbf{P}-box} \times 10^{-11}$} \\
\midrule
$\pi^+$ & $0.67(1)$ & $-15.6(0.2)$ \\
$K^+$ & $0.59(1)$ & $-0.48(0.02)$ \\
$\pi_1^+$ & $0.58(1)$ & $-0.0020(1)$ \\
\bottomrule
\end{tabularx}
\end{table}

\subsection{Axial-Vector Mesons}
While pseudoscalar mesons provide the dominant component of the HLbL amplitude, other hadronic degrees of freedom can also contribute. Among these, axial-vector mesons have received particular attention. These are a key source of the current uncertainty\,\cite{Aliberti:2025beg} and are essential to enforce short-distance constraints\,\cite{Melnikov:2003xd,Ludtke:2024ase,Bijnens:2019ghy,Bijnens:2020xnl,Bijnens:2021jqo}, and numerous subtleties arise in this sector (see, e.g.,\,\cite{Roig:2019reh,Mager:2025pvz,Cappiello:2025fyf}).%

The study of axial-vector mesons remains particularly challenging within the DSE framework. Although they may, at a simplistic level, be viewed as orbital excitations of the pseudoscalar ground states, their proper description already requires interaction kernels that go beyond the standard implementations\,\cite{Xu:2022kng,Qin:2020jig,Qin:2020rad,Xu:2026sjy}. This difficulty is manifest even at the level of the mass spectrum, while developing a rigorous, symmetry-preserving framework for computing their electromagnetic elastic and transition FFs presents an even greater challenge\,\cite{Xing:2025iab,Eichmann:2024glq}. An additional complication stems from the need to characterize \mbox{a rather} large tower of excited states\,\cite{Roig:2019reh,Mager:2025pvz,Cappiello:2025fyf}, whose treatment remains highly nontrivial within the DSE formalism, e.g.,\,\cite{Miramontes:2021exi,Chang:2019eob,Li:2016dzv}.

Despite these limitations, the axial-vector contribution has been explored at different levels within the DSE framework. First, Ref.\,\cite{Eichmann:2024glq} addressed the problem within \mbox{a computationally} involved setup that, under a number of considerations, permits the use of the standard RL truncation. For the axial-vector multiplet, $\mathrm{AV}=\{a_1,f_1,f_1'\}$, \mbox{this yields}
\begin{equation}
a_\mu^{\mathrm{AV}}=17.4(5.7)\times10^{-11},
\,
a_\mu^{\mathrm{AV\text{-}tower}}=24.8(6.1)\times10^{-11},
\end{equation}
where the first result includes only the lowest-lying multiplet, and the second incorporates \mbox{a larger} tower of axial-vector excitations. The errors reported in Ref.\,\cite{Eichmann:2024glq} have been combined in quadrature.

More recently, Ref.\,\cite{Xing:2025iab} explored the same contribution within a numerically simpler, yet conceptually rich framework that provides greater insight into the role of symmetries and the underlying mathematical structure. The resulting estimate is more exploratory in nature and yields
\begin{equation}
    a_\mu^{\mathrm{AV}}=8.16_{-1.02}^{+1.22}\times 10^{-11}\,.
\end{equation}

With the above in mind, the next step is therefore to develop a symmetry-preserving treatment beyond RL that retains the realistic momentum dependence while incorporating the structures required by the axial-vector sector.

\subsection{Proton-Box Contribution}
The proton is already accessible within the DSE formalism, as its electromagnetic structure has been investigated using some of the most sophisticated DSE methods available (e.g.,\,\cite{Yao:2024uej,Eichmann:2016yit,Qin:2019hgk}). These studies are consistent with the explorations of the mesonic sector presented herein. Moreover, the DSE framework gives access to the flavor-separated electromagnetic FFs, opening the possibility of investigating isospin-symmetry-breaking effects\,\cite{Miramontes:2022mex,Miramontes:2026ffk,Bashir:2026keh}.

Fermion-box formulae are also available\,\cite{Colangelo:2017fiz,Colangelo:2019uex}, and a first lattice-based exploration has already been conducted\,\cite{Estrada:2024rfi}. Following the procedure described therein and adopting the DSE data from Ref.\,\cite{Yao:2024uej}, we obtain
\begin{equation}
\label{eq:protonbox}
    a_\mu^{p-\text{box}}=1.74(4)\times10^{-12},
\end{equation}
where the quoted uncertainty accounts for the uncertainties in the model parameters (see Ref.\,\cite{Yao:2024uej}) and in the Padé$_{1,3}$ interpolant used to represent the numerical data. {This constitutes the first DSE-based assessment of this contribution.}

{Early estimates of this contribution based on the Heavy Mass Expansion\,\cite{Kuhn:2003pu} place this quantity at the level of $a_\mu^{p-\textrm{box}}=9.7\times10^{-11}$, which would make it comparable in magnitude to the contributions discussed above. Nevertheless, as discussed in Ref.\,\cite{Estrada:2024rfi}, its magnitude is not controlled solely by the proton mass, but also by the momentum dependence of its electromagnetic FF. Employing lattice QCD inputs, the authors of that study find
\begin{equation}
    a_\mu^{p-\text{box}}=1.82(7)\times10^{-12}.
\end{equation}

In this context, the agreement between our first DSE-based evaluation and the \mbox{first lattice QCD} determination\,\cite{Estrada:2024rfi} is particularly significant, as both approaches account for the momentum dependence of the proton electromagnetic FFs. Furthermore, despite the relative suppression, the resulting proton-box contribution is comparable in magnitude to the current theoretical uncertainty in $a_\mu$ and may therefore become increasingly relevant as the precision of the Standard Model prediction improves.
}

\textls[-15]{Interestingly, although the DSE calculation of the proton FFs employs equal current quark masses $m_u=m_d$, the interactions generate a nontrivial partition of the total result, with the $u$-quark contribution accounting for roughly $63\%$ and the $d$-quark  for the \mbox{remaining 37\%:}}
\begin{equation}
    a_\mu^{p_u-\text{box}}=1.10(3)\times10^{-12}\,,\,a_\mu^{p_d-\text{box}}=0.64(2)\times10^{-12}.
\end{equation}

Evidently, the flavor contributions do not simply follow the naive $2/3$--$1/3$ partition expected from the proton's valence-quark content. Rather, they reflect the nontrivial dynamics encoded in its internal structure\,\cite{Lu:2022cjx,Wang:2024fjt,Yao:2024uej}. 

{The proton-box contribution illustrates how the dynamics of emergent hadronic structure can feed into precision observables, while its nontrivial flavor decomposition provides a window into the underlying quark-level dynamics. This highlights the potential of the DSE framework to resolve hadronic contributions at the flavor level and, ultimately, to quantify the impact of isospin-symmetry breaking.

Although the proton is expected to set the overall scale of the baryonic contribution, once $a_\mu^{p\text{-box}}$ is well determined, studying the remaining members of the baryon octet would allow us to quantify its flavor dependence and assess the effects of $SU(3)$-flavor-symmetry breaking. Extending the analysis to the spin-$3/2$ decuplet would, in turn, provide \mbox{an opportunity} to explore the role of spin and higher electromagnetic multipoles. However, the latter extension is considerably more challenging, as a general formalism for these contributions has yet to be developed. Moreover, there is no guarantee that their inclusion would lead to a meaningful reduction in the uncertainty at the current level of precision, or even in the foreseeable future.}

{
\section{Perspectives on HVP and DSEs}
\label{sec:HVP}
The HVP contribution constitutes the largest source of uncertainty in the SM prediction of $a_\mu$\,\cite{Aliberti:2025beg}. Its leading-order contribution can be determined primarily through dispersive methods and lQCD. Nevertheless, despite substantial progress in both, important challenges remain, including tensions among experimental datasets in the dispersive approach and systematic uncertainties in lattice QCD, including isospin-breaking (IB) effects\,\cite{Aliberti:2025beg}. \mbox{An additional} avenue for constraining the dominant HVP contribution is provided by the hadronic $\tau^-\to\pi^-\pi^0\nu_\tau$ spectrum, provided that IB effects are properly accounted for when the spectrum is related to the corresponding $e^+e^-$ cross section\,\cite{Flores-Baez:2025nra,Perez:2026yxg}.

Early evaluations of the HVP contribution within the DSE framework\,\cite{Goecke:2011pe} demonstrated the potential of this approach to address this problem at the quark level. A more recent assessment\,\cite{Miramontes:2026ffk} has highlighted the importance of quantifying the effects of both strong and electromagnetic IB. In particular, it suggests a net IB-induced shift of
\begin{equation}
    \Delta a_\mu^{\text{HVP,LO}}=4.5\times 10^{-10}. 
\end{equation}

Flavor sensitivity can also be investigated through strong decays accessible within the DSE framework, e.g.,\,\cite{Bashir:2026keh,Xu:2024rew,daSilveira:2022pte}. In particular, IB in the $\rho\pi\pi$ coupling was recently investigated for the first time within a symmetry-preserving DSE framework\,\cite{Bashir:2026keh}, yielding an IB correction to the leading-order HVP contribution of
\begin{equation}
    \Delta a_\mu^{\text{HVP,LO}}[\rho\pi\pi]=-2.32(1.64)\times 10^{-10}.
\end{equation}

\textls[-15]{Further studies of processes such as $K^*\to K\pi$ and $\phi\to K\bar K$ may provide additional insights into IB effects and their implications for the hadronic contributions to $a_\mu$. More broadly, the systematic improvability of the DSE framework offers a path toward an increasingly complete description of these contributions, rooted in the underlying \mbox{quark-level dynamics.}}

Although relatively modest, IB effects are clearly not negligible at the level of precision required for $a_\mu$. Their systematic treatment may therefore help reduce the uncertainty in the HVP contribution and, ultimately, improve the precision of the SM prediction.
}
\section{Summary and Outlook}
\label{sec:Summary}

We have discussed general aspects of the Dyson–Schwinger equations formalism and revisited a selection of results obtained within this framework concerning the hadronic light-by-light contributions to the muon anomalous magnetic moment. Specifically, we have revisited the pseudoscalar-pole and pseudoscalar-box contributions, the axial-vector meson case, and the very first proton-box exploration within DSEs. {We have also provided some perspectives on the application of this formalism to the hadronic vacuum polarization piece.}

Over the past two decades, the DSE framework has developed to a stage at which a broad range of non-perturbative quantities associated with QCD and hadron structure are within reach. Numerous form factors with electromagnetic, electroweak and gravitational probes, as well as a wide variety of structure functions, have been explored within this approach. The resulting predictions invariably show consistency with both robust experimental measurements and theoretical expectations.

These developments have progressively established the DSE framework as a powerful and versatile tool for the calculation of hadronic contributions to precision observables, such as $a_\mu$. The results presented here, and in particular their agreement with data-driven determinations, provide further evidence for the reliability of the DSE approach in estimating these quantities. More broadly, the present results highlight the capacity of the DSE formalism to probe the non-perturbative QCD dynamics responsible for the emergence of hadron structure.

\textls[-15]{Beyond the specific topics discussed in this document, the systematic development of the DSE framework opens the possibility of examining an increasingly broad range of hadronic observables within a single unified continuum-QCD description. This is particularly timely given the stringent demands on the theoretical control of the hadronic components of $a_\mu$.}

The results presented herein point toward a broader program of systematically improvable DSE calculations of the hadronic contributions to $a_\mu$. The axial-vector meson case can be extended to increasingly sophisticated treatments and, within the same framework, to scalar and tensor mesons. Further baryonic contributions and hadronic processes remain to be explored. In particular, {for the hadronic vacuum polarization component,} a systematic study of strong decays in the light-quark sector could provide a means of quantifying isospin-symmetry-breaking effects and, in turn, assessing their impact on the uncertainties of the hadronic contributions to $a_\mu$.

\vspace{6pt}

\funding{This work was supported by the Spanish Ministry of Science and Innovation (MICINN grant nos. PID2022-140440NB-C22, PID2025-171916NB-C32) and Junta de Andalucía (grant no. P18-FR-5057).
}

\dataavailability{No new data were generated  in this study. All experimental data discussed in this work were obtained from previously published sources, which are appropriately cited throughout the manuscript.} 

\acknowledgments{K.R.
 gratefully acknowledges A. Bashir, P. Roig and J. Rodríguez-Quintero
 for their invaluable support and encouragement throughout the development of his scientific career. The author also acknowledges the Physics Department at CINVESTAV-Zacatenco for its hospitality and support.}

\conflictsofinterest{The author declares no conflicts of interest.} 



\abbreviations{Abbreviations}{%
The following abbreviations are used in this manuscript:
\\

\noindent 
\begin{tabular}{@{}ll}
BRL & Beyond Rainbow-Ladder \\
BS & Bethe–Salpeter \\
DSE & Dyson–Schwinger equations \\
EHM & Emergence of Hadron Masses \\
FF & Form Factor \\
HVP & Hadronic Vacuum Polarization\\
HLbL & Hadronic Light-by-Light\\
IB & Isospin Breaking\\
lQCD & Lattice Quantum Chromodynamics \\
PTIR & Perturbation Theory Integral Representation \\
PDG & Particle Data Group \\
QCD & Quantum Chromodynamics \\
QGV & Quark–gluon vertex \\
QPV & Quark–photon vertex \\
RL & Rainbow-ladder \\
SM & Standard Model\\
\end{tabular}
}

{
\appendixstart
\appendix
\section{Integral Representations}
\label{app:appendix}
To parametrize the quark propagator beyond the domain covered by the numerical solutions, it is convenient to express it in terms of complex-conjugate poles. With flavor indices omitted, it reads: 
	\vspace{-3pt}
	\begin{eqnarray}
	S(p)&=& -i\gamma \cdot p\;\sigma_v(p^2)+\sigma_s(p^2)=\sum^{j_m}_{j=1}\left[\frac{z_j}{i\gamma{\cdot}p+m_j}+\frac{z^{*}_j}{i\gamma{\cdot}p+m^{*}_j}\right],\label{quarkPTIR}
	\end{eqnarray}
	where $z_j,m_j$ are obtained from a best fit to the numerical solutions, ensuring \textrm{Im}$(m_j)\neq 0\forall j$, a feature consistent with confinement. We find that $j_m=2$ is adequate to provide \mbox{an accurate interpolation.}

    As established before (see Equation\,\eqref{eq:BSEps}),  the BS amplitude of a pseudoscalar meson is written in terms of four Dirac covariants $\mathcal{F}=E_{\textbf{P}},F_{\textbf{P}},G_{\textbf{P}},H_{\textbf{P}}$.	Each of these scalar functions is split in two parts and can be parametrized in terms of PTIRs in the following way:
	\begin{subequations}\label{BSAPTIR}
		\begin{eqnarray}
		\mathcal{F}(k;P)&&=\mathcal{F}^i(k;P)+\mathcal{F}^u(k;P)\,,\\
		\mathcal{F}^i(k,P)&&=c^i_{\mathcal{F}}\int^1_{-1}dz\rho_{\nu^i_{\mathcal{F}}}(z)\big[a_{\mathcal{F}}\widehat{\Delta}^4_{\Lambda^i_{\mathcal{F}}}(k^2_z)+a^-_{\mathcal{F}}\widehat{\Delta}^5_{\Lambda^i_{\mathcal{F}}}(k^2_z)\big]\,,\\
		\mathcal{F}^u(k;P)&&=c^u_{\mathcal{F}}\int^1_{-1}dz\rho_{\nu^u_{\mathcal{F}}}(z)\widehat{\Delta}^{{l^u_{\mathcal{F}}}}_{\Lambda^u_{\mathcal{F}}}(k^2_z)\,.
		\end{eqnarray}
	\end{subequations}
	Here, $\widehat{\Delta}_{\Lambda}(s)=\Lambda^2\Delta_\Lambda(s)$, $k^2_z=k^2+zk{\cdot}P$, and $a^-_\mathcal{F}=1-a_\mathcal{F}$. The indices `$i$' and `$u$' denote the connection with the infrared and ultraviolet behaviors of the BS amplitude. For practical purposes, the spectral density can be expressed as
	\begin{equation}
	\rho_\nu(z)=\frac{\Gamma\left[\frac{3}{2}+\nu\right]}{\sqrt{\pi}\Gamma\left[1+\nu\right]}(1-z^2)^\nu\,,
	\end{equation}
	where the interpolating parameters are obtained through fitting to the Chebyshev moments:
	\begin{align}
	\mathcal{F}_n(k^2)=\frac{2}{\pi}\int^1_{-1}dx\sqrt{1-x^2}\mathcal{F}(k;P)U_n(x)\,.
	\end{align}
	
	In our calculations, $n=0,2$, where $U_n$ is an order-$n$ Chebyshev polynomial of the second kind. Furthermore, since $\mathcal{F}_4(k;P)$ is small and has no impact, it is omitted in all cases. For similar reasons, $\mathcal{F}_3(k;P)$ might be omitted for $\eta'$ and $\eta_b$ as well.
	
	The quadratic forms adopted for the propagators and BS amplitudes, together with a suitable representation of the quark–photon interaction, allow us to employ analytical techniques such as Feynman parametrization. This makes it possible to access large photon virtualities in form-factor calculations by replacing the numerical integration over momenta with integrations over Feynman parameters and the spectral density. 
    
    The specific parametrizations of the propagators and BS amplitudes introduced in Equations~(A1) and (A2a--c)
 have proven to be highly useful, with their corresponding interpolation parameters reported in Refs.~\cite{Raya:2015gva,Raya:2016yuj,Ding:2018xwy}. Alternative representations have been also employed in the computation of light-front wave functions, e.g.,\,\cite{Raya:2021zrz,Raya:2026gwg}.

}


\begin{adjustwidth}{-\extralength}{0cm}
\reftitle{References}


\bibliography{main}

\PublishersNote{}
\end{adjustwidth}
\end{document}